\documentclass[11pt,a4paper]{article}

\usepackage{jheppub,tcolorbox}
\usepackage[T1]{fontenc}
\usepackage{amsmath,amssymb}
\usepackage{tikz}
\usepackage{tikz-cd}
\usepackage{ctable}
\usepackage[normalem]{ulem}

\usetikzlibrary {decorations.pathmorphing}
\usetikzlibrary {decorations.markings}
\usetikzlibrary {positioning,shapes.misc}
\usetikzlibrary {decorations.shapes,shapes.geometric} 
\usetikzlibrary{graphs}
\usetikzlibrary{shadings}

\tikzset{
  paint/.style={draw=#1!50!black, fill=#1!50},
  my star/.style={decorate,decoration={shape backgrounds,shape=star}, star points=#1}
  }

\title{\boldmath{Iterative Solution of the Kerr Black Hole Metric}}

\author[a,b]{Poul H. Damgaard}
\author[c]{Hojin Lee}
\author[d]{Kanghoon Lee}
\author[d]{Tabasum Rahnuma}

\affiliation[a]{Niels Bohr International Academy, Niels Bohr Institute, University of Copenhagen, Blegdamsvej 17, DK-2100 Copenhagen \O, Denmark}
\affiliation[b]{Theoretical Physics Department, CERN, 1211 Geneva 23, Switzerland}
\affiliation[c]{School of Physics, Korea Institute for Advanced Study, 85 Hoegi-ro, Dongdaemun-Gu, Seoul 02455, Korea}
\affiliation[d]{Quantum Universe Center, Korea Institute for Advanced Study, 85 Hoegi-ro, Dongdaemun-Gu, Seoul 02455, Korea}

\emailAdd{phdamg@nbi.dk}
\emailAdd{hojinlee@kias.re.kr}
\emailAdd{kanghoon.lee1@gmail.com}
\emailAdd{tabasum03@kias.re.kr}

\abstract{Using a recursive solution of the Einstein equations, we consider the perturbative expansion of the metric corresponding to a Kerr black hole. Because the metric is a function of two parameters, Newton's constant G and the Kerr spin parameter a, the perturbation theory naturally becomes a double expansion. In harmonic gauge the recursion relations can be solved to arbitrarily high orders in these two expansion parameters but to re-sum the series into the closed-form harmonic gauge metric requires the introduction of terms that are redundant and correspond to the addition of harmonic functions to the coordinates. Issues related to dimensional regularization of Fourier transforms are explained in detail.}

\begin{document}

\preprint{CERN-TH-2026-105}
\maketitle

\section{Introduction}

In contrast to the perturbative expansions of quantum mechanics and quantum field theory, which are known to be only asymptotic in general, there is a tantalizing possibility that post-Minkowskian expansions of classical generel relativity may be convergent in specific physical situations. A recursive formalism was used in ref. \cite{Damgaard:2024fqj} to first solve and subsequently sum the perturbative series of a Schwarzschild black hole. A natural question that arises is whether this can be
generalized to other exact solutions of the Einstein field equations. The case of a Kerr black hole may be the first that comes to mind, 
and in this paper we explore the extent to which a similar all-order calculation of the Kerr metric can be carried through, and
summed within the expected range of convergence.

The possibility of finding convergent all-order perturbative solutions to the Einstein field equations is of both conceptual and practical significance. Although one should not expect to find convergence of the post-Minkowskian expansion for arbitrary boundary conditions, it is particularly important to understand if the perturbative expansion corresponding to the post-Minkowskian two-body problem is convergent or not. Superficial analogies with relativistic quantum field theories might suggest otherwise. But it should be kept in mind that the fact that perturbative expansions of quantum field theories are only asymptotic is intimately related to quantum mechanical effects such as tunneling or the appearance of thresholds, $i.e.$, massive or massless loops where pairs can go on-shell and thus lead to particle-antiparticle production. Although, as is well known, there can be great advantages to solving the classical field equations of Einstein gravity by means of quantum field theoretic methods, these amplitude-based \cite{Bjerrum-Bohr:2018xdl,Cheung:2018wkq,Kosower:2018adc,Bern:2019nnu,Bern:2019crd,Cristofoli:2019neg,Damgaard:2019lfh,Bjerrum-Bohr:2019kec,DiVecchia:2020ymx,DiVecchia:2023frv,Bjerrum-Bohr:2021vuf,Bjerrum-Bohr:2021din,Damgaard:2021ipf,Cho:2021nim,Lee:2022aiu,Damgaard:2023ttc,Cho:2023kux,Bern:2020buy,Bern:2022kto,Damgaard:2022jem,Brandhuber:2022enp,Adamo:2022ooq,Bjerrum-Bohr:2023jau,Aoude:2023dui,Bjerrum-Bohr:2025lpw,Akpinar:2025tct,Brunello:2025eso} or worldline-based techniques \cite{Goldberger:2004jt,Levi:2014gsa,Kalin:2020mvi,Mogull:2020sak,Jakobsen:2021smu,Dlapa:2021npj,Jakobsen:2022psy,Kalin:2022hph,Jakobsen:2023ndj,Damgaard:2023vnx,Ben-Shahar:2023djm} of course discard all such quantum effects. This suggests that the question is wide open and it is therefore important to settle the issue. Exploring the perturbative expansion of general relativity in the neighborhood of compact matter sources can give insights into this question, and adding spin to the matter source gives us one new territory in which to explore the potential convergence of the post-Minkowskian expansion.

Convergence or divergence of the post-Minkowskian expansion should preferably be phrased in terms of physical observables such as the scattering angle (or bending of light) from flat space to flat space, the periastron precession, or a gravitational wave observed far from the source. Nevertheless, it is difficult to envisage how one might be able to reconcile any finite and convergent prediction for such observables without the corresponding metric of just one single localized source itself also being represented in terms of a convergent series from perturbation theory. We are thus impelled to first try to understand the isuue of convergence of the perturbative series in the simplest possible settings of static or stationary solutions for the metric in general relativity.

There are three ingredients that greatly facilitate the perturbative solution of the Einstein equations of motion. The first is the use of Landau-Lifshitz (or gothic) variables when expanding around flat Minkowski space. The second is the use of harmonic gauge which is ideal for the Green function method in general relativity. The third is perhaps the most crucial for large-order calculations: the introduction of auxiliary field variables that furbish the inverse of the Landau-Lifshitz metric density variables but whose components are treated as independent until the end of the calculation. The latter allows for a compact expression of the equations of motion which avoids the infinite series normally associated with the inversion of the metric. 

The weak-field perturbative expansion of the metric would ideally be in terms of Newton's constant in appropriate dimensionless units. In the case of the Schwarzschild metric the expansion is thus in terms of $GM/r$, where $M$ is the mass and $r$ is the distance from the origin. For the perturbative expansion of the Kerr metric there are two length scales: distance $r$ and $a \equiv J/M$, where $J$ is angular momentum. Because distance $r$ initially is taken large, a small parameter in which to expand is still $GM/r$, but it is then less clear which counting scheme should be used for $a$. An additional natural dimensionless parameter to expand in could be $a/r$. Such a perturbative counting scheme corresponds to that of the multipole expansion of the metric outside a general compact object \cite{Damgaard:2026kqg}. A double expansion does not preclude a summation of the perturbation series, and the known Kerr metric in harmonic coordinates does indeed suggest, as will be explained in detail in the next section, that it should be summable in terms of geometric series in suitable combinations of $r, a$ and $G$. The first approach to the perturbative metric based on Feynman diagrams is that of Duff \cite{Duff:1973zz}; see also refs. \cite{Bjerrum-Bohr:2002fji,Gambino:2024uge,Scheopner:2023rzp,Bianchi:2024shc} for the inclusion of (quantum) spin.

The iterative solution for the Kerr metric starts with identifying a suitable source. This will yield the solution for the metric at the linearized level in the $G$-expansion. In contrast to amplitude-based or worldline-based methods, the subsequent steps are algebraic: the next order in $G$ follows by gluing the leading-order contributions together, and the structure iterates to any order in $G$. This method is a return to basics, the solution of the Einstein field equations without any allusion to quantum field theory, scattering amplitudes or worldline diagrams. The reason this method is so attractive just now is that it builds on the large progress made recently based on the synergy with those quantum field theory methods. In particular, the techniques for carrying out what in quantum field theory language are loop integrals (arising from our viewpoint from gluing lower order terms together through the recursion) have developed significantly in recent years. We recover similar integrals starting directly with the equations of motion but it is natural to organize the integrations differently. This new point of view suggests that also the two-body problem in general relativity with advantage can be attacked by the same recursive formalism.  

Our iterative approach is formulated in momentum space: the recursion relations follow from the equations of motion using Fourier transforms of the currents. In addition to providing powerful loop integral machinery, another advantage of working in momentum space is that the nontrivial causal structure, which is difficult to handle in position space, can be implemented systematically through use of the retarded Green function. This will become important when solving time-dependent problems. However, the Fourier transforms generally generate divergences at intermediate steps, and an appropriate regularization is thus required in order to properly define the currents in momentum space. We adopt dimensional regularization, which is the most natural and efficient regularization scheme in perturbative gauge theories. In the classical general relativity literature, dimensional regularization has mostly been studied from the perspective of position space. Here we instead examine it from the viewpoint of momentum space. 

An essential issue that is often glossed over in dimensional regularization is that certain algebraic identities in an integer number of dimensions do not continue to hold after dimensional regularization. As an example, consider the relation $\delta r^2 \equiv r^2 - x^2 - y^2 - z^2 = 0$ (or its momentum-space counterpart $\delta\ell^2 \equiv |\boldsymbol{\ell}|^2 - \ell_x^2 - \ell_y^2 - \ell_z^2 = 0$) which is satisfied only in three dimensions, and not for $d = 3 - 2\epsilon$. Perhaps surprisingly, even if a function $f(\delta r^2)$ vanishes in the limit $\epsilon \to 0$, its Fourier transform need not vanish in that limit, $\mathcal{F}[f(\delta r^2)]\big|_{\epsilon \to 0} \neq 0$,  in general. The same issue clearly arises for inverse Fourier transforms. Such observations should not affect the final outcome (in the above example one does indeed recover the starting point upon Fourier transform followed by its inverse, no matter which prescription one chooses). 

To address these issues systematically, we thus first need to establish that the convolution theorem for Fourier transformations remains valid in dimensional regularization. We then use this to show that solving the equations of motion in momentum space is equivalent to solving them in position space, irrespective of prescriptions.  We solve the recursion relations iteratively and explicitly construct the currents and corresponding metric perturbations up to fourth post-Minkowskian order and to all orders in the spin parameter $a$. We highlight the gauge prescription that gives agreement with the Kerr metric in harmonic coordinates as derived in ref. \cite{Lin:2014laa}.

This paper goes through all steps to set up and carry out the procedure outlined above. We start in Section 2 with presenting the exact solution of the Kerr metric in harmonic coordinates together with a source for the metric. In Section 3 we provide the perturbative Einstein equations, and in Section 4 we derive a recursion relation from these equations of motion. We discuss the various issues with dimensional regularization in detail and propose a prescription that handles them consistently. Equally crucial (and not unrelated) is the appearance of gauge redundancies that reveal how the very special form of the compact form of the exact Kerr metric is achieved through an implicit use of this gauge freedom. In section 5, we explicitly solve the recursion and derive all orders in $a$ up to 4th order in $G$. Finally, Section 6 contains our conclusions.


\section{Kerr Metric in Harmonic Coordinates}

We begin by recalling the standard representation of the Kerr metric in Boyer-Lindquist coordinates $\bar{x}^{\mu} = \{\bar{t},\bar{r},\bar{\theta},\bar{\phi}\}$
\begin{align}
\begin{split}
    ds^2 & =-\left(1-\frac{2GM \bar{r}}{\Sigma} \right)d\bar{t}^2 + \frac{\Sigma}{\Delta}d\bar{r}^2 + \Sigma d\bar{\theta}^2
    \\
    & \quad +\left( \bar{r}^2 + a^2  + \frac{2GM \bar{r} a^2}{\Sigma} \sin^2 \bar{\theta} \right) \sin^2 \bar{\theta} d \bar{\phi}^2 -\frac{4GM \bar{r} a\sin^2\bar{\theta}}{\Sigma} d\bar{t} d\bar{\phi} \,,
\end{split}\label{Kerr_BL}
\end{align}
where $\Sigma = \bar{r}^2+a^2 \cos^2 \bar{\theta}$ and $\Delta = \bar{r}^2- 2GM \bar{r} +a^2$. In these coordinates, the outer and inner horizons are located at $\bar r_{\pm}=GM \pm \sqrt{G^{2}M^{2}-a^{2}}$\,, where $\bar r_{+}$ and $\bar r_{-}$ correspond to the outer and inner horizon, respectively.  Moreover, the locus $\bar r=0$ is not a point: it describes a disk of radius $a$. 
The curvature singularity lies on the boundary of this disk, forming a ring characterized by
\begin{equation}
\bar r=0\,,\qquad \bar{\theta}=\frac{\pi}{2}\,.
\end{equation}


We now consider harmonic coordinates of the Kerr metric. We can derive such harmonic coordinates $X_{\mu}(\bar{x}^{\mu})$, where $\bar{x}^{\mu} = \{\bar{t},\bar{r},\bar{\theta},\bar{\phi}\}$, by solving the defining equation for a given metric $g_{\mu\nu}$
\begin{equation}
  \Box_{g} X_{\mu}(\bar{x}^{\mu}) = g^{\rho \sigma} \frac{\partial^{2} X_{\mu}}{\partial \bar{x}^{\rho} \partial \bar{x}^{\sigma}}-g^{\rho \sigma} \Gamma_{\rho \sigma}^{\lambda} \frac{\partial X_{\mu}}{\partial \bar{x}^{\lambda}} = 0\,,
\label{}\end{equation}
with the condition
\begin{equation}
  g^{\mu\nu} \Gamma_{\mu\nu}^{\rho} = \partial_{\mu}\big(\sqrt{-g} g^{\mu\rho}\big) = 0\,.
\label{}\end{equation}
and it justifies the name harmonic.
Solving for the condition, and relabelling $X_{\mu}$ as $X_{0} = t$, $X_{1} = x$, $X_{2} = y$ and $X_{3} = z$, we obtain the following relation between Boyer-Lindquist coordinates and harmonic coordinates:
\begin{equation}
  \bar{t}=t\,,
  \qquad
  x+ i y = \sqrt{R^2+a^2} \sin \bar{\theta} \, e^{i \Phi} \,, 
  \qquad 
  z= R \cos \bar{\theta} \,,
\end{equation}
where
\begin{equation}
\begin{aligned}
  R = \bar{r} - GM \,, 
  \qquad
  \Phi = \bar{\phi} + \int \frac{a}{\Delta} d\bar{r} + \tan^{-1}\frac{a}{R} \,,
\end{aligned}\label{harmonic_coordinates}
\end{equation}
and $R$ satisfies
\begin{equation}
  R^{2} = \frac{1}{2}\big(r^2-a^2+\sqrt{(r^2-a^2)^2+4 a^2 z^2}\big)\,,
  \qquad
  r^{2} = x^{2}+y^{2}+z^{2}\,.
\label{def_R}\end{equation}
It is important to note that the harmonic coordinates $X^{\mu}$ do not cover the entire space, but only the region where $\bar{r}>GM$, as follows from $\bar{r} = R + GM$. The corresponding Kerr metric in these coordinates takes the form \cite{Lin:2014laa}
\begin{equation}
\begin{aligned} 
  d s^{2} 
  =& 
  - d t^{2}+\frac{R^{2}(R+GM)^{2}+a^{2} z^{2}}{\left(R^{2}+\frac{a^{2}}{R^{2}} z^{2}\right)^{2}}
  \left[ \big(R^{2}+a^{2}-G^{2}M^{2}\big) A^{2} 
  +\frac{z^{2}B^{2}}{R^{2}\big(R^{2}-z^{2}\big)} \right] 
  \\& 
  +\frac{2 GM(R+GM)}{(R+GM)^{2}+\frac{a^{2}}{R^{2}} z^{2}}\left[
      d t 
    + \frac{R G^{2}M^{2} a^{2}\left(R^{2}-z^{2}\right)A}{\left(R^{2}+a^{2}\right)\left(R^{4}+a^{2} z^{2}\right)}
    + \frac{a\left(y d x-x d y\right)}{R^{2}+a^{2}}\right]^{2} 
  \\ & 
  +\frac{(R+G^{2}M^{2})^{2}+a^{2}}{R^{2}-z^{2}}\left[\frac{G^{2}M^{2} aR^{2} \left(R^{2}-z^{2}\right)A}{\left(R^{2}+a^{2}\right)\left(R^{4}+a^{2} z^{2}\right)}
  	+ \frac{a\left(y d x-x d y\right)}{R^{2}+a^{2}}
  \right]^{2}\,,
\end{aligned}\label{KerrHarmonicCoordinate}
\end{equation}
where the 1-form fields $A$ and $B$ are defined by
\begin{equation}
  A = \big(R^{2}+a^{2}-G^{2}M^{2}\big)^{-1}\Big(\mathbf{x} \cdot d \mathbf{x}+\frac{a^{2}}{R^{2}} z d z\Big) \,,
  \qquad 
  B = \mathbf{x} \cdot d \mathbf{x}-R^{2}z^{-1} d z \,.
\label{}\end{equation}

As mentioned in the Introduction, we will not work with the metric $g_{\mu\nu}$ itself, but with the Landau-Lifshits variables defined by
\begin{equation}
  \mathfrak{g}^{\mu\nu} = \sqrt{-g} g^{\mu\nu}\,, 
  \qquad
  \mathfrak{g}_{\mu\nu} = \frac{1}{\sqrt{-g}} g_{\mu\nu}\,. 
\label{Landau_Lifshitz}\end{equation}
In these so-called `gothic' variables, the perturbation of the Schwarzschild metric away from the Minkowski metric takes a remarkably simple form \cite{Damgaard:2024fqj}. As will be demonstrated, the Kerr metric exhibits a similar property. Moreover, the harmonic condition also admits a concise expression as
\begin{equation}
  \partial_{\mu} \mathfrak{g}^{\mu\nu} = 0\,.
\label{harmonic_condition}\end{equation}

The Kerr metric in harmonic coordinates \eqref{KerrHarmonicCoordinate} in terms of these Landau-Lifshitz variables is given by
\begin{equation}
\begin{aligned}
  \mathfrak{g}^{00}
  &=
  -\frac{R^2 (G M+R) \left(a^2 (3 G M+R)+(G M+R)^3\right)}{\big(a^2-G^2 M^2+R^2\big)\big(a^2 z^2+R^4\big)}+\frac{a^2 z^2}{a^2 z^2+R^4}\,,
  \\
  \mathfrak{g}^{01}
  &=
  \frac{2 a G M R^2 y (G M+R)}{\left(a^2 z^2+R^4\right) \left(a^2-G^2 M^2+R^2\right)}\,,
  \qquad
  \mathfrak{g}^{02}
  =
  -\frac{2 a G M R^2 x (G M+R)}{\left(a^2 z^2+R^4\right) \left(a^2-G^2 M^2+R^2\right)}\,,
  \\
  \mathfrak{g}^{03}
  &=
  0\,,
  \qquad
  \mathfrak{g}^{11}  
  =
  1-\frac{G^2 M^2 R^2 (a y+R x)^2}{\left(a^2+R^2\right)^2 \left(a^2 z^2+R^4\right)}\,,
  \qquad
  \mathfrak{g}^{12}
  =
  \frac{G^2 M^2 R^2 (a y+R x) (a x-R y)}{\left(a^2+R^2\right)^2 \left(a^2 z^2+R^4\right)}\,,
  \\
  \mathfrak{g}^{13}
  &=
  -\frac{G^2 M^2 R z (a y+R x)}{\left(a^2+R^2\right) \left(a^2 z^2+R^4\right)}\,,
  \qquad
  \mathfrak{g}^{22}
  =
  1-\frac{G^2 M^2 R^2 (a x-R y)^2}{\left(a^2+R^2\right)^2 \left(a^2 z^2+R^4\right)}\,,
  \\
  \mathfrak{g}^{23}
  &=
  \frac{G^2 M^2 R z (a x-R y)}{\left(a^2+R^2\right) \left(a^2 z^2+R^4\right)}\,,
  \qquad
  \mathfrak{g}^{33}
  =
  1-\frac{G^{2} M^{2}z^2 x}{a^2 z^2+R^4}\,.
\end{aligned}\label{LandauLifshitzExpression}
\end{equation}

We now define perturbations $\mathfrak{h}^{\mu \nu}$ of $\mathfrak{g}^{\mu\nu}$ around a flat Minkowskian background in the weak-field regime
\begin{equation}
  \mathfrak{g}^{\mu \nu}
  =
  \eta^{\mu \nu}-\mathfrak{h}^{\mu \nu}\,. 
\label{}\end{equation}
For deriving the perturbative solution of the Kerr metric in \eqref{LandauLifshitzExpression}, the most natural framework is the post-Minkowskian (PM) expansion, a power-series expansion in Newton's constant $G$:
\begin{equation}
  \mathfrak{h}^{\mu\nu} = \sum_{n=1}^{\infty} G^{n} \mathfrak{h}^{\mu\nu}\big|_{n}\,.
\label{}\end{equation}
However, the PM expansion alone still yields expressions that depend on the non-polynomial function of $r$ which is included in the function 
$R(x,y,z)$ of \eqref{def_R}, as well as the asymmetric treatment of coordinates due to the spin parameter $a$. As an example, the non-vanishing components of the leading-order $\mathfrak{h}^{\mu\nu}$ is given by
\begin{equation} \label{Potential_Expansion}
  \mathfrak{h}^{00}\big|_{1} =\frac{4M R^3}{R^4+a^2 z^2}\,,
  \qquad
  \mathfrak{h}^{0\alpha}\big|_{1} = - \frac{2a M R^3 \epsilon^{\alpha\beta}x_{\beta}}{(R^2+a^2)(R^4+a^2z^2)} \,.
\end{equation}
Thus unless the solution is further organized as a power series in $1/r$, explicit computations become cumbersome. For this reason, we also perform a series expansion in the small spin parameter $a$, and hence consider a double expansion in both $G$ and $a$, 
\begin{equation}
  \mathfrak{h}^{\mu\nu}
  =
  \sum_{m=1}^{\infty} \sum_{n=0}^{\infty} G^{m} a^{n}\mathfrak{h}^{\mu\nu} \big|_{m,n}\,.
\label{}\end{equation}
Expanding in this manner, we find the non-vanishing components of $h^{\mu\nu}_{(1)}$ at linear order in $G$ to be given by
\begin{equation}
\begin{aligned}
  \mathfrak{h}^{00}\big|_{1,2n}
  &=
  4M \frac{(-1)^n}{r^{2n+1}}\,P_{2n}\left(\frac{z}{r}\right)\,.
  \\
  \mathfrak{h}^{01}\big|_{1,2n+1}
  &= 
  (-1)^n \frac{2My}{(2n+1)r^{2n+2}\sqrt{x^{2}+y^{2}}}P^1_{2n+1}\left(\frac{z}{r} \right)
  \\
  \mathfrak{h}^{02}\big|_{1,2n+1}
  &=
  (-1)^{n+1} \frac{2Mx}{(2n+1)r^{2n+2}\sqrt{x^{2}+y^{2}}} P^1_{2n+1}\left(\frac{z}{r} \right)
  \\
  \mathfrak{h}^{03}\big|_{1,n}
  &= 0  \,,
\end{aligned}\label{1PM_h}
\end{equation}
where $P_{k}(x)$ and $P_{\ell}^{m}(x)$ are Legendre polynomials and associated Legendre polynomials, respectively.

As is well known, a general multipole expansion of the metric favors a formalism described in terms of so-called Symmetric and Trace-Free (STF) tensors. The axial symmetry of Kerr restricts each rank-$\ell$ STF multipole tensor to be proportional to the axis-aligned STF tensor $\hat{s}^L$ (where $L$ is a multi-index of length $\ell$) built from the unit spin vector. This follows from the fact that imposing invariance under the axial subgroup $SO(2)_s \subset SO(3)$ selects a one-dimensional invariant subspace within each rank-$\ell$ irreducible representation, so that the only admissible direction is that of the symmetry axis (see Appendix C of Ref.~\cite{Damgaard:2026kqg}). To determine the proportionality coefficient, we expand the 1PM Kerr metric in powers of $1/r$, thus obtaining a series in Legendre polynomials $P_\ell(\hat{n}\cdot\hat{s})$. Using the identity relating Legendre polynomials to STF products,
\begin{equation}
    P_\ell(\hat{n}\cdot\hat{s}) = \frac{(2\ell-1)!!}{\ell!}\,\hat{n}^L\hat{s}^L,
\end{equation}
we rewrite this expansion in the STF basis and compare term by term with the general 1PM multipole expansion of Ref.~\cite{Damgaard:2026kqg}. Reading off the coefficients, the Kerr mass and current multipoles are then
\begin{equation}
    M^{L} = (-1)^{\ell}\,\mathrm{M}_{2\ell}\,\hat{s}^L, \qquad S^{aL} = (-1)^{\ell}\,\mathrm{M}_{2\ell+1}\,\hat{s}^{aL},
\end{equation}
where $\mathrm{M}_{\ell} = M a^{\ell}$.

When comparing our iterative solution for the Kerr metric with the closed-form harmonic-gauge Kerr metric of ref.~\cite{Lin:2014laa}, an interesting difference occurs already at order $G^2$. Specifically, for the components $\mathfrak{h}^{ij}|_{2}$ we find a term in the Kerr metric of ref.~\cite{Lin:2014laa} at order $G^2$ which is proportional to $a$, 
\begin{equation}
    \mathfrak{h}^{ij}\big|_{2,1} = 2 \frac{x^k x^{(i}\epsilon^{j)k3}}{r^5} ~.
\end{equation}
In the language of general multipoles this would have to arise from a mass monopole times current dipole ($MS_i$) combination. However, as will become clear below, our recursive equations do not generate this term. The resolution, which has already been discussed in ref.~\cite{Damgaard:2026kqg} is that harmonic gauge does not completely fix the coordinates: the harmonic gauge condition which leads to $\square\,\xi^i = 0$ allows residual coordinate shifts by any harmonic vector $\xi^i $. The closed-form Kerr metric of ref.~\cite{Lin:2014laa} implicitly uses one particular choice within this residual freedom. At order $G^2$ and to linear order in $a$, we find such a gauge vector $\xi^i|_{\mathcal{O}(a^{1})}= -M^2\,a \,\epsilon^{ij3}\frac{x^j}{3r^3}$ or, generalizing, a gauge vector to all odd powers of $a$,
\begin{equation}
    \xi^i = \frac{M^2}{2a^2}\left[\frac{Ra}{R^2+a^2} + \arctan\!\left(\frac{R}{a}\right) - \frac{\pi}{2}\right]\epsilon^{ij3}x^j ~.
\label{2PMgauge}\end{equation}
This satisfies $\square\,\xi^i = 0$ and it transforms $\mathfrak{h}^{ij}|_{2}$ so as to remove or add the offending terms that differ. The parts in $\mathfrak{h}^{ij}|_{2}$ that are odd in $a$ are therefore gauge artefacts within the harmonic gauge:  the closed-form expression of eq. \eqref{LandauLifshitzExpression} will be related to our iterative solution by this residual gauge transformation.


\subsection{Source of the Kerr black hole in harmonic gauge}\label{sec:source}

We now turn to the derivation of a source for a Kerr metric. Israel was the first to show \cite{Israel:1970kp} that a source for Kerr can be understood as a composite object consisting of a disk and a ring, where the ring appears as a regulator localized on the boundary of the disk. We shall here determine the source using a new inverse method that starts from a certain linearized field rather than postulating the source a priori. The resulting source agrees precisely with the result derived from the multipole expansion in \cite{Bianchi:2024shc}. Since we work with the Landau--Lifshitz variables, which are tensor densities, it is natural to formulate the source in terms of a density $\mathfrak{j}^{\mu\nu}$ rather than the usual energy--momentum tensor. 

Our analysis is based on solving Einstein equations in the post-Minkowskian expansion. Once the gravitational field at leading order is specified, all higher-order PM fields shall be fixed through the nonlinear structure of the Einstein equations. If we assume a stationary source with no dependence on Newton's constant $G$, the field at linearized level is determined solely by the external source $\mathfrak{j}^{\mu\nu}$ via the linearized Einstein equations. To this order, they are a set of Poisson equations providing the relation between the field and the source. Since the explicit form of $\mathfrak{h}^{\mu\nu}|_{1}$ is already given by the Kerr metric discussed above, we can invert this relation and reconstruct a corresponding source. This inverse procedure allows us to identify a source for the Kerr metric in harmonic coordinates, including both the disk and the regulating ring at its boundary.

At leading order in $G$, the field equations reduce to Poisson equations in harmonic gauge, 
\begin{equation}
  \square \mathfrak{h}^{\mu\nu}\big|_{1,n} = - 2 \mathfrak{j}^{\mu\nu}\big|_{1,n}\,.
\label{Poisson}\end{equation}
The purely spatial components of the metric perturbation at this order are trivial, $\mathfrak{h}^{ij}|_{1}=0$, which implies that the corresponding source components satisfy $\mathfrak{j}^{ij} = 0$. We therefore decompose the source into two independent sectors and derive $\mathfrak{j}^{00}$ and $\mathfrak{j}^{0i}$ separately.

\subsubsection{The source $\mathfrak{j}^{00}$}
The stationary $00$-component $\mathfrak{h}^{00}|_{1}$ can be viewed as a Newtonian potential generated by a mass distribution $\mathfrak{j}^{00}$. As we have seen above, $\mathfrak{h}^{00}|_{1}$ for the Kerr metric in harmonic coordinates can be represented by even Legendre polynomials \eqref{Potential_Expansion}. This form already suggests that the effective source is axisymmetric, confined to the equatorial plane, and carries an infinite tower of even-parity multipole moments. In particular, the absence of odd-$\ell$ terms implies reflection symmetry $z\to -z$ and rules out any net mass dipole moment. Furthermore, one can show that $\nabla^{2}\mathfrak{h}^{00}|_{1, 2n}=0$ except at $z=0$, and the convergence radius of the series expansion \eqref{1PM_h} is $r>a$. It is therefore natural to model the source as a thin disk of radius $a$ in the plane $z=0$. 

We now use the standard Newtonian (or, by analogy, electrostatic) solution for the potential generated by a thin axisymmetric disk to reconstruct the mass density. In this case, the source term is given by $\mathfrak{j}^{00} = 8\pi G \sigma(\rho)\,\delta(z)\,\Theta(a-\rho)$. Let us take a surface mass density $\sigma(\rho)$ supported on $z=0$ and $0\le\rho\le a$, where $\rho=\sqrt{x^{2}+y^{2}}$. The corresponding gravitational potential is given by
\begin{equation}
  \mathfrak{h}^{00}\big|_{1}
  =
  \frac{1}{2\pi} \int d^3\vec r'\, \frac{\mathfrak{j}^{00}}{|\vec{r}-\vec{r}^{\,\prime}|}
  =
  \int d^3\vec r\,'\, \frac{4\sigma(\rho')\,\delta(z')\,\Theta(a-\rho')}{|\vec r-\vec r\,'|}\,.
\label{GeneralDisk}\end{equation}
Expanding the Green function in spherical harmonics and performing the trivial azimuthal integration $\phi$, one finds the well-known Legendre expansion for an axisymmetric disk \cite{Damgaard:2026kqg}
\begin{equation}
  \mathfrak{h}^{00}\big|_{1} (r,\theta)
  =
  8\pi \sum_{n=0}^{\infty} \frac{(-1)^n \Gamma\!\big(n+\frac{1}{2}\big)}{n!\sqrt{\pi}}\,\frac{P_{2n}(\cos\theta)}{r^{2n+1}} \int_0^a d\rho'\,\rho'^{2n+1}\,\sigma(\rho')\,.
\label{GeneralDiskLegendre}\end{equation}
Next, we determine the surface mass density $\sigma(\rho)$. By comparing \eqref{GeneralDiskLegendre} and $\mathfrak{h}^{00}|_{1,2n}$ in \eqref{1PM_h} at $z=0$ plane, the integral on the right-hand side of \eqref{GeneralDisk} should be identified as
\begin{equation}
  \int_0^a d\rho'\,\rho'^{2n+1}\,\sigma(\rho')
  =
  \frac{M a^{2n} n!\sqrt{\pi}}{2\pi \Gamma\!\big(n+\frac{1}{2}\big)}\,.
\label{sourceMoment}\end{equation}
However, the mass density $\sigma(\rho)$ cannot be determined from this equation alone. Motivated by the fact that in electromagnetism the charge density on a disk or annulus is typically expressed as a function of $\sqrt{a^2 - \rho^2}$, we adopt the following {\em ansatz}:
\begin{equation} 
  \sigma(\rho) = \frac{A}{(a^2 - \rho^2)^{\frac{m}{2}}}\,, 
  \qquad
  m\in \mathbb{Z}_{+}\,,
\label{}\end{equation}
where $A$ is a constant to be determined later.
This density diverges in the limit $\rho \to a$, so we introduce an infinitesimal regulator $\varepsilon$ through 
$(a^{2}-\rho^{2})^{-\frac{m}{2}} \to (a^{2}+\varepsilon^{2}-\rho^{2})^{-\frac{m}{2}}$. 

We now substitute this ansatz into the integral above to determine the value of $m$ that yields the desired form. To carry this out efficiently, we define the following two-parameter integral 
\begin{eqnarray}
  I_{n,m} &\equiv& \int_0^{a} d \rho \frac{\rho^{2n+1}}{(a^2+\varepsilon^{2}-\rho^2)^{\frac{m}{2}}} \cr
  &=&
  \frac{(a^{2}+\varepsilon^{2})^{n+1-\frac{m}{2}}}{2} B_{\frac{a^{2}}{a^{2}+\varepsilon^{2}}}\left(n+1,1-\frac{m}{2}\right) ,
\end{eqnarray}
where $B_{z}(a,b)$ is the incomplete beta function. 
Expanding with respect to the regulator $\varepsilon$ yields
\begin{equation}
  I_{n,m}
  =
  \frac{a^{2 n-m+2}}{2} \frac{n!\Gamma\left(1-\frac{m}{2}\right)}{\Gamma\left(n-\frac{m}{2}+2\right)}
  -\frac{1}{2}\!\! \sum_{\substack{p=0 \\ 2 p \leq m-2}}^{n} \!\!\!\binom{n}{p}\frac{p! \Gamma(1-\frac{m}{2})}{\Gamma(p+2-\frac{m}{2})}a^{2 n-2 p} \varepsilon^{2 p+2-m}
  + \mathcal{O}(\varepsilon)\,.
\label{expansionImn}\end{equation}
Note that if $m$ is odd, the $\varepsilon^{0}$ order piece arises only from the first term, and the second term is divergent as $\varepsilon\to 0$ in powers of $1/\varepsilon$. However, if $m$ is even, the second term can also have $\varepsilon^{0}$ terms depending on $m$ and $p$, but the gamma functions can be singular, generating a $\log \varepsilon$ term in their product. Thus we consider only the case of $m$ being odd.

Comparing the finite part of \eqref{expansionImn} and $\mathfrak{h}^{00}|_{1}$ in \eqref{sourceMoment}, this fixes $m$ to take the value $m=3$ and also the overall constant $A$. The corresponding mass density $\sigma(\rho)$ is then
\begin{equation}
  \sigma(\rho) 
  =
  -\frac{Ma}{2\pi(a^2+ \varepsilon^{2}-\rho^2)^{3/2}}\,,\qquad 0\le \rho\le a\,.
\end{equation}
In this case, $I_{n,3}$ reduces to
\begin{equation}
  I_{n,3}
  =
    \frac{a^{2n}}{\varepsilon} 
  - \frac{a^{2n-1} n! \sqrt{\pi}}{\Gamma \left( n+\frac{1}{2} \right)} 
  + \mathcal{O}(\varepsilon) \,.
\label{MomentIntegral}\end{equation}
However, the total mass of this disk diverges as $\epsilon \to 0$
\begin{equation}
  M_{\rm disk} = 2\pi \int_0^a d\rho\,\rho\,\sigma(\rho)
  =
  - M a I_{0,3}
  =
  - \frac{M a}{\varepsilon} + M \,.
\end{equation}
and the singular term of \eqref{GeneralDiskLegendre} is found from the expansion \eqref{expansionImn}, which gives
\begin{equation}
  - \frac{4M}{\varepsilon} \sum_{n=0}^{\infty} \frac{(-1)^n \Gamma\!\big(n+\frac{1}{2}\big)}{n!\sqrt{\pi}}\,\frac{a^{2n+1}}{r^{2n+1}} P_{2n}(\cos\theta)\,.
\label{}\end{equation}

A simple way to remove the $1/\epsilon$ divergence is to supplement the disk by a ring source of radius $a$ and with a line density of mass $\lambda$. For an axisymmetric ring located at $\rho=a$, $z=0$, the contribution to the potential can also be expanded in even Legendre polynomials. The potential from the ring is given by
\begin{equation} \label{Ring_Potential}
  \lambda \sum_{n=0}^{\infty}\frac{(-1)^n}{n!} \frac{\Gamma \left( n+\frac{1}{2} \right)}{\sqrt{\pi}} \left( \frac{a}{r}\right)^{2n+1} P_{2n}(\cos \theta) \,.
\end{equation}
Thus, if we set $\lambda=\frac{4 M}{\epsilon}$, this cancels exactly the divergent term arising from the regulated disk. The combined potential of the regulated disk and the ring masses is finite in the limit $\epsilon\to 0$ and gives rise to the metric components $\mathfrak{h}^{00}_{(1)}$ of eq. \eqref{1PM_h}.

Combining the regulated disk plus ring configuration, we obtain
\begin{equation}
  \mathfrak{j}^{00} (\vec r)
  = 4 GM\lim_{\varepsilon\to 0} \left[
  	\frac{1}{\varepsilon}\,\delta(a-\rho)
	- \frac{a}{\big(a^2+\varepsilon^2-\rho^2\big)^{3/2}}\,\Theta(a-\rho)
  \right]\delta(z)\,.
\end{equation}
This result agrees with \eqref{Bianchi:2024shc}, which is based on multipole expansion. Furthermore, from this we obtain the total mass
\begin{equation}
  \int d^3\vec r\,\sigma(\rho) = M\,.
\end{equation}
The source thus has a clear geometric interpretation: it is a thin disk of radius $a$ with a singular mass density regularized by a compensating ring at its boundary. The disk contributes a negative effective mass density, which is partially cancelled by the positive ring so that the net mass is $M$. Although such a distribution is not physically realistic as a matter stress-energy tensor, it is perfectly acceptable as an effective source that reproduces the Kerr geometry in harmonic gauge and it will serve as the starting point for our higher-order iterative construction in the following.

We finally note that in momentum space the corresponding source $\mathfrak{j}^{00}$ is given by the simple expression \cite{Bianchi:2024shc}
\begin{equation}
  \mathfrak{j}^{00} = 8\pi GM \int_{\vec{k}} e^{i\vec{k}\cdot \mathbf{x}} \cos( a\, k_{\perp})\,, 
  \qquad
  k_{\perp}=\sqrt{k_{x}^{2}+k_{y}^{2}}\,.
\label{00sourceMomentum}\end{equation}

\subsubsection{The source $\mathfrak{j}^{0i}$}
\label{subsec:disk-ring-current}

We now turn to the determination of the external source $\mathfrak{j}^{0i}$. Since this corresponds to the mass current of the system, it is natural to interpret it as the current generated by a rigid rotation of the composite disk and ring mass-distribution identified above. This construction is directly analogous to stationary electrodynamics. The mass current $\mathfrak{j}^{0i}$ plays a role analogous to that of a stationary electric current density, serving as the source term for the vector potential in the field equations.

For a stationary axisymmetric rotating solution with spin aligned along the $z$-axis, this $\mathfrak{j}^{0i}$ field is proportional to the azimuthal unit vector $\hat\phi^i$ due to rotational symmetry. In cylindrical coordinates $(\rho,\phi,z)$, we therefore take it as a surface current supported on the equatorial plane:
\begin{equation}
  \mathfrak{j}^{\phi} \hat{\phi} 
  = 
  8\pi G K(\rho)\,\delta(z)\Theta(a-\rho)\,\hat\phi\,,
  \qquad
  \hat\phi = (-\sin\phi,\cos\phi,0)\,,
\end{equation}
where $K(\rho)$ is the surface current density.
Thus $\mathfrak{h}^{0i}|_{1}$ is indeed proportional to $\hat{\phi}^{i}$, and we correspondingly consider the angular $h^{\phi}$ component 
\begin{equation}
  \mathfrak{h}^{\phi}\big|_{1}
  = -\sin\phi \mathfrak{h}^{01}\big|_{1} + \cos\phi \mathfrak{h}^{02}\big|_{1}
  =
  \sum_{n=0}^{\infty} (-1)^{n+1} \frac{2Ma^{2n+1}}{(2n+1)r^{2n+2}} P^1_{2n+1}\left(\cos \theta \right)\,.
\label{}\end{equation}

From the current source,
\begin{equation}
\begin{aligned}
  \mathfrak{h}^{\phi}\big|_{1}  &=
  \int \rho' d \rho^{\prime} d \phi^{\prime}\frac{4K(\rho')\Theta(a-\rho')\delta(z')}{\left|\vec{r}-\vec{r}'\right|}\hat{\phi} \cdot \hat{\phi}'\,.
\end{aligned}\label{}
\end{equation}
Again, we expand the Green function and perform the integral over the azimuthal angle. In contrast to the case of scalar potential in \eqref{GeneralDiskLegendre}, the $\phi'$ integral is not trivial in this case due to $\hat{\phi} \cdot \hat{\phi'} = \cos(\phi-\phi')$, and we obtain
\begin{equation}
  \mathfrak{h}^{\phi}\big|_{1}  =
  -4\pi \sum_{n=0}^{\infty} \frac{(-1)^{n}\Gamma(n+1/2)}{(n+1)!\sqrt{\pi}} \frac{P_{2 n+1}^{1}(\cos \theta)}{r^{2 n+2}} 
  \int_{0}^{a} K(\rho) \rho^{2 n+2} d \rho\,. 
\label{h0i_1PM_Expansion}\end{equation}
Since we are considering a rotating disk with charge density $\sigma(\rho)$, we find that $K(\rho)$ must be given by
\begin{equation}
  K(\rho) = -\frac{M \rho}{4\pi\big(a^2+\varepsilon^2-\rho^2\big)^{3/2}} \,.
\end{equation}
Indeed, when we evaluate the moment integral $I_{n+1,3}$ using the formula of eq. \eqref{MomentIntegral},
\begin{equation}
  I_{n+1,3}
  =
    \int_{0}^{a} d\rho\, \frac{ \rho^{2n+3}}{\big(a^2+\varepsilon^2-\rho^2\big)^{3/2}} 
  =
    \frac{a^{2n+2}}{\varepsilon}
  - \frac{a^{2n+1}(n+1)! \sqrt{\pi}}{\Gamma \left( n+\frac{3}{2} \right)}
  + \mathcal{O}(\varepsilon) \,,
\end{equation}
we see that the non-singular part of \eqref{h0i_1PM_Expansion} exactly reproduces eq. \eqref{h0i_1PM_Expansion}, and the singular part is 
\begin{equation}
  h^{\phi}_{(1),{\rm singular}}
  = \frac{M}{\varepsilon} \sum_{n=0}^{\infty} a^{2n+2}\frac{(-1)^{n}\Gamma(n+1/2)}{(n+1)!\sqrt{\pi}} \frac{P_{2 n+1}^{1}(\cos \theta)}{r^{2 n+2}}\,.
\label{}\end{equation}
%

%
%
The combined regularized current profile in Cartesian coordinates is thus 
\begin{equation}
\begin{aligned}
  \mathfrak{j}^{01}(\rho)
  &=
  -2GM\,\lim_{\epsilon\to 0}\left[
    \frac{1}{\epsilon}\,\delta(a-\rho)
    -\frac{\rho}{\big(a^2+\epsilon^2-\rho^2\big)^{3/2}}\,
    \Theta(a-\rho) \right]\delta(z)\frac{y}{\rho}\,,
  \\
  \mathfrak{j}^{02}(\rho)
  &=
  2GM\,\lim_{\epsilon\to 0}\left[
    \frac{1}{\epsilon}\,\delta(a-\rho)
    -\frac{\rho}{\big(a^2+\epsilon^2-\rho^2\big)^{3/2}}\,
    \Theta(a-\rho) \right]\delta(z)\frac{x}{\rho}\,,
  \\
  \mathfrak{j}^{03}(\rho)
  &=
  0\,.
\end{aligned}\label{}
\end{equation}
Again, this agrees with \cite{Bianchi:2024shc}. We note that also their Fourier transforms become remarkably simple, and we have 
\begin{equation}
\begin{aligned}
  \mathfrak{j}^{01}(\rho)
  &=
  i4\pi GM \int_{\vec{k}} \sin \left(a\, k_{\perp}\right) \frac{k_{y}}{k_{\perp}}\,,
  \\
  \mathfrak{j}^{02}(\rho)
  &=
  -i4\pi GM \int_{\vec{k}} \sin \left(a\, k_{\perp}\right) \frac{k_{x}}{k_{\perp}}\,,
  \\
  \mathfrak{j}^{03}(\rho)
  &=
  0\,,
\end{aligned}\label{0isourceMomentum}
\end{equation}
closely mirroring the case of $\mathfrak{j}^{00}$.

\subsubsection{Interpretation}

\begin{figure}[t]
  \centering
  \includegraphics[width=\textwidth]{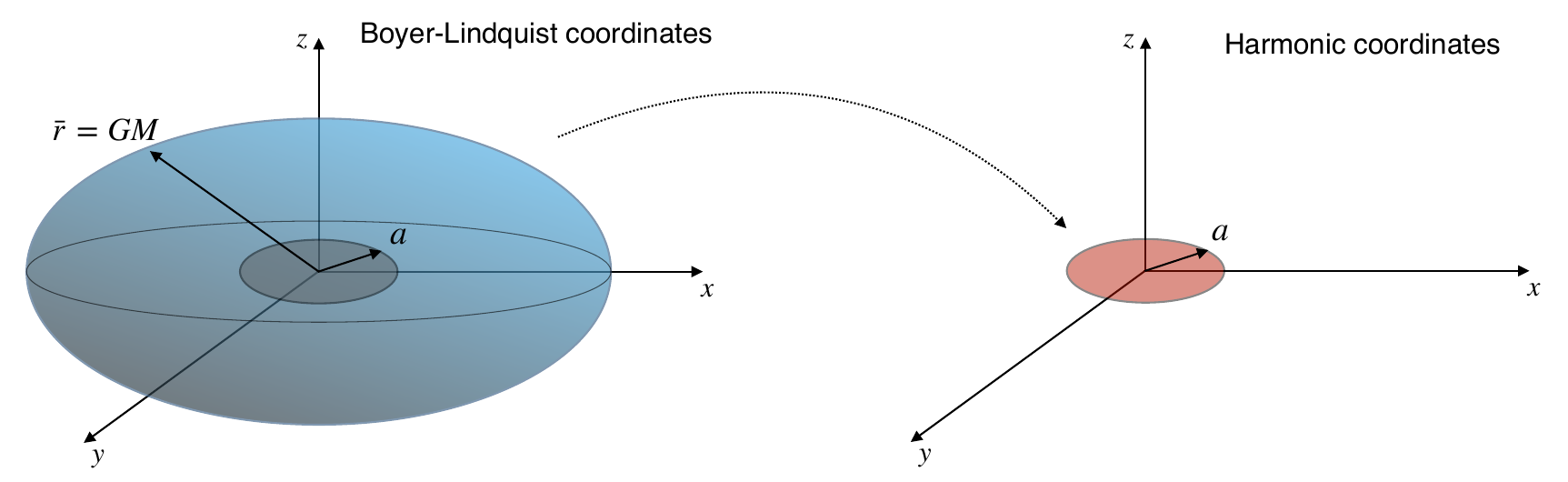}
  \caption{}
\label{fig1}\end{figure}

In harmonic coordinates, the locus $R=0$ is mapped to a disk of radius $a$. This region corresponds to the domain defined by $\bar{r} \leq GM$ in Boyer-Lindquist coordinates as we defined in \eqref{harmonic_coordinates}, which represents the region exterior to the physical source. In particular, the Boyer-Lindquist surface $\bar r=GM$ should be identified with an oblate spheroid, which is mapped by the coordinate transformation to the surface $R=0$, see Figure 1. As a result, the set $R=0$ in harmonic coordinates corresponds precisely to the disk of radius $a$.

This implies that the source inferred from the harmonic gauge is an \emph{effective} source that reproduces the exterior Kerr geometry in the domain of validity of our construction. Thus it can be understood as a boundary condition corresponding to the harmonic coordinates of a spinning black hole. The actual physical source generating the Kerr metric is located in the interior region, i.e.\ inside the disk bounded by $R=0$.

\section{Solving the Einstein Equations Perturbatively} \label{Sec:3}
In this section we derive the explicit form of the perturbative Einstein equations in a form which is useful for our purposes. We will use these perturbative equations to construct the post-Minkowskian expansion of the Kerr black hole solution in harmonic coordinates through the source described in the previous section. The perturbative expansion of the Einstein equations is notoriously cumbersome in the conventional framework of metric perturbation theory based on the expansion for the metric as $g_{\mu\nu} = \eta_{\mu\nu}+h_{\mu\nu}$. In fact, the origin of such a complexity is not difficult to identify, and it is two-fold: the appearance of $\sqrt{-g}$ (which must be expanded in powers of $h$) and the need to also involve the inverse metric $g^{-1}$ (which leads to a geometric series in $h$). 

Our strategy is to remove both obstacles from the beginning. The first is quite standard:to employ the Landau--Lifshitz variables introduced above. This helps by absorbing factors of $\sqrt{-g}$ and is justified as a more economical way of parametrizing metric perturbations. Second, we proceed by treating the metric density and its inverse (in Landau--Lifshitz variables) as independent dynamical variables at intermediate stages. Their relationship is recovered by imposing the inverse identity as a constraint. Rather than substituting a formal series expansion for the inverse field, we solve the inverse constraint directly and substitute the resulting expression back into the equations of motion. With this idea, the perturbative field equations take a remarkably simple form, and the pseudo energy-momentum tensor truncates at finite order in $\mathfrak{h}$.


Consider the Einstein-Hilbert action $S_{\rm EH}$ coupled to an external source $\mathfrak{j}^{\mu\nu}$ in terms of the Landau-Lifshitz variables defined in eq. \eqref{Landau_Lifshitz},
\begin{equation}
  S[\mathfrak{g}^{\mu\nu},\mathfrak{j}^{\mu\nu}] 
  =
  S_{\rm EH}[\mathfrak{g}^{\mu\nu}]
  + \frac{1}{16\pi G} \int \mathrm{d}^{4}x \frac{1}{\sqrt{-\mathfrak{g}}}\mathfrak{g}_{\mu\nu} \mathfrak{j}^{\mu\nu}\,.
\label{total_action}\end{equation}
The Einstein-Hilbert action $S_{\rm EH}[\mathfrak{g}]$ can be written in terms of the metric density $\mathfrak{g}$ and its inverse:
\begin{equation}
\begin{aligned} 
  S_{\mathrm{EH}}[\mathfrak{g}^{\mu\nu}]
  = \frac{1}{16\pi G}\int \mathrm{d}^{D} x \bigg[ & 
  	  \frac{1}{4} \mathfrak{g}^{\mu \nu} \partial_{\mu} \mathfrak{g}^{\rho \sigma} \partial_{\nu} \mathfrak{g}_{\rho \sigma}
  	- \frac{1}{2} \mathfrak{g}^{\mu \nu} \partial_{\mu} \mathfrak{g}^{\rho \sigma} \partial_{\rho} \mathfrak{g}_{\nu \sigma}+(D-2) \mathfrak{g}^{\mu \nu} \partial_{\mu} d \partial_{\nu} d 
  \\&
  	+\partial_{\mu}\left(2 \mathfrak{g}^{\mu \nu} \partial_{\nu} d-\partial_{\nu} \mathfrak{g}^{\mu \nu}\right)\bigg]\,,
\end{aligned}
\label{EH_action}\end{equation}
where $\partial_{\mu} d = -\frac{1}{2(D-2)} \mathfrak{g}_{\rho \sigma} \partial_{\mu} \mathfrak{g}^{\rho \sigma}$. We see that the use of Landau--Lifshitz variables has made all explicit factors of $\sqrt{-g}$ disappear. 

For further simplification, and following the method employed in the Schwarzschild case \cite{Damgaard:2024fqj}, we treat $\mathfrak{g}^{\mu\nu}$ and $\mathfrak{g}_{\mu\nu}$ on equal footing by introducing a new field $\tilde{\mathfrak{g}}_{\mu\nu}$, which is forced to equal the gothic inverse $\mathfrak{g}_{\mu\nu}$, by imposing this relation as a constraint:
\begin{equation}
  \mathfrak{g}^{\mu\rho}\tilde{\mathfrak{g}}_{\rho\nu} = \delta^{\mu}{}_{\nu}\,.
\label{Inverse_relation_constraint}\end{equation}
We introduce independent linearized perturbations of $\mathfrak{g}^{\mu\nu}$ and $\tilde{\mathfrak{g}}_{\mu\nu}$ around the flat Minkowskian backround metric $\eta$
\begin{equation}
  \mathfrak{g}^{\mu\nu} = \eta^{\mu\nu} - \mathfrak{h}^{\mu\nu} \,,
  \qquad
  \tilde{\mathfrak{g}}_{\mu\nu} = \eta_{\mu\nu} + \tilde{\mathfrak{h}}_{\mu\nu} \,.
\label{metric_perturbation}\end{equation}
It follows that  $\mathfrak{h}^{\mu\nu}$ and $\tilde{\mathfrak{h}}_{\mu\nu}$ are related through \eqref{Inverse_relation_constraint}, which gives
\begin{equation}
  \tilde{\mathfrak{h}}^{\mu\nu} = \mathfrak{h}^{\mu\nu} + \mathfrak{h}^{\mu\rho} \tilde{\mathfrak{h}}_{\rho}{}^{\nu}\,,
\label{}\end{equation}
where indices are raised and lowered by the flat Minkowski metric. 

Therefore, in what follows we first derive the Einstein equations of motion in the usual way by varying the action $S_{\rm EH}[\mathfrak{g}]$ in eq. \eqref{EH_action} with respect to the gothic metric $\mathfrak{g}^{\mu\nu}$. The resulting field equations are expressed in terms of $\mathfrak{g}^{\mu\nu}$ and its inverse $\mathfrak{g}_{\mu\nu}$ as usual. Next, we replace every occurrence of $\mathfrak{g}_{\mu\nu}$ by $\tilde{\mathfrak{g}}_{\mu\nu}$. This prescription is equivalent to starting from an action written in terms of $\tilde{\mathfrak{g}}_{\mu\nu}$ (instead of $\mathfrak{g}_{\mu\nu}$) and adding a Lagrange multiplier that enforces the constraint of eq. \eqref{Inverse_relation_constraint}, $i.e.$,
\begin{equation}
\label{eq:tildeg-LM-action}
  S_{\rm EH}\big[\mathfrak{g}\big]
  \to 
  \tilde{S}_{\rm EH}\big[\mathfrak{g}^{\mu\nu},\tilde{\mathfrak{g}}_{\mu\nu}\big]
  +
  \frac{1}{16\pi G} \int d^{4}x\, \lambda^{\mu}{}_{\nu}\,
  \Big(\mathfrak{g}^{\nu\rho}\tilde{\mathfrak{g}}_{\rho\mu}-\delta^{\nu}{}_{\mu}\Big)\,.
\end{equation}
The corresponding field equations will of course agree exactly with those obtained by the replacement rule above.

Let us now consider variations of the action $S[\mathfrak{g},\mathfrak{j}]$ in eq. \eqref{total_action} with respect to $\mathfrak{g}_{\mu\nu}$
\begin{equation}
\begin{aligned}
  \frac{\delta S[\mathfrak{g}^{\mu\nu},\mathfrak{j}^{\mu\nu}]}{\delta\mathfrak{g}_{\mu\nu}}
  &=
  - \frac{1}{16\pi G} \mathcal{G}^{\mu\nu} 
  + \frac{1}{16\pi G} \frac{1}{\sqrt{-\mathfrak{g}}} \bigg(\mathfrak{j}^{\mu\nu} -\frac{1}{2} \mathfrak{g}^{\mu\nu}\mathfrak{j}^{\rho\sigma} \mathfrak{g}_{\rho\sigma}\bigg) \,,
\end{aligned}\label{}
\end{equation}
where $\mathcal{G}_{\mu\nu}$ is the Einstein tensor density defined by $\mathcal{G}^{\mu\nu} = -\frac{\delta S_{\rm EH}}{\delta \mathfrak{g}_{\mu\nu}}$ \footnote{Note that the determinant in $\sqrt{-\mathfrak{g}}$ is defined by the lower indices object, $\mathfrak{g}= \det \mathfrak{g}_{\mu\nu}$, and its variation is given by the Jacobi formula of variations,
\begin{equation*}
  \delta \frac{1}{\sqrt{-\mathfrak{g}}} 
  = -\frac{1}{2} (-\mathfrak{g})^{-\frac{3}{2}} \big(- \delta \mathfrak{g}\big) 
  = -\frac{1}{2} (-\mathfrak{g})^{-\frac{3}{2}} \big(- \mathfrak{g} \delta\mathfrak{g}_{\mu\nu} \mathfrak{g}^{\mu\nu}\big)
  = -\frac{1}{2} \frac{1}{\sqrt{-\mathfrak{g}}} \delta\mathfrak{g}_{\mu\nu} \mathfrak{g}^{\mu\nu}\,.
\label{}\end{equation*}
}
where the Einstein tensor density which we replaced $\mathfrak{g}_{\mu\nu}$ to $\tilde{\mathfrak{g}}_{\mu\nu}$
\begin{equation}
\begin{aligned} 
  \mathcal{G}^{\mu\nu}= &\
    \frac{1}{2} \partial_{\kappa} \big(\mathfrak{g}^{\kappa\lambda} \partial_{\lambda} \mathfrak{g}^{\mu\nu}\big) 
  - \frac{1}{2} \partial_{\lambda} \mathfrak{g}^{\kappa\mu} \partial_{\kappa} \mathfrak{g}^{\lambda\nu}
  - \mathfrak{g}^{\kappa(\mu} \partial_{\kappa} \partial_{\lambda} \mathfrak{g}^{\nu) \lambda}
  + (D-2) d^{\mu} d^{\nu}
  + \mathfrak{g}^{\mu\nu} \partial_{\kappa} d^{\kappa}
  \\&
  +\frac{1}{2} \mathfrak{g}^{\kappa\alpha} \mathfrak{g}^{\beta(\nu}\partial_{\alpha}\mathfrak{g}^{\mu)\lambda} \partial_{\kappa}\tilde{\mathfrak{g}}_{\lambda\beta}
  - \mathfrak{g}^{(\mu|\alpha|} \mathfrak{g}^{\nu)\beta}\partial_{\alpha}\mathfrak{g}^{\kappa\lambda} \partial_{\kappa}\tilde{\mathfrak{g}}_{\lambda\beta}
  +\frac{1}{4} \mathfrak{g}^{\mu\alpha} \mathfrak{g}^{\nu\beta}\partial_{\alpha}\mathfrak{g}^{\kappa\lambda} \partial_{\beta}\tilde{\mathfrak{g}}_{\kappa\lambda}
  \,,
\end{aligned}
\label{EinsteinTensorDensity}\end{equation}
where
\begin{equation}
  d^{\mu} \equiv - \frac{1}{2(D-2)} \mathfrak{g}^{\mu\kappa} \tilde{\mathfrak{g}}_{\rho \sigma}\partial_{\kappa} \mathfrak{g}^{\rho \sigma} \,.
\label{}\end{equation}
Then the Einstein equations of motion with the external source $\mathfrak{j}^{\mu\nu}$ are given by
\begin{equation}
  \mathcal{G}^{\mu\nu} 
  = 
  \frac{1}{\sqrt{-\tilde{\mathfrak{g}}}}
  \bigg( 
    	\mathfrak{j}^{\mu\nu}
  	  - \frac{1}{2} \mathfrak{g}^{\mu\nu} \mathfrak{j}^{\rho\sigma} \mathfrak{g}_{\rho\sigma}
  \bigg) \,.
\label{}\end{equation}
where $\sqrt{-\tilde{\mathfrak{g}}}=\sqrt{-\det \tilde{\mathfrak{g}}_{\mu\nu}}$ .

If we expand $\mathcal{G}^{\mu\nu}$ using the definition of metric perturbations of eq.  \eqref{metric_perturbation} in harmonic gauge, $\partial_{\mu} \mathfrak{h}^{\mu\nu} = 0$, the linear order in the perturbation field is given by
\begin{equation}
\begin{aligned}
  \mathcal{G}^{\mu\nu}
  &=
  - \frac{1}{2}\Box \mathfrak{h}^{\mu\nu} +\frac{1}{4} \eta^{\mu\nu} \Box \mathfrak{h}
  + \mathcal{O}(\mathfrak{h}^{2})\,,
  \qquad
  \mathfrak{h} = \mathfrak{h}^{\rho}{}_{\rho}
  \\
  &= K^{\mu\nu}{}_{\rho\sigma} \mathfrak{h}^{\rho\sigma}  + \mathcal{O}(\mathfrak{h}^{2})\,,
\end{aligned}\label{Gupup}
\end{equation}
where $K^{\mu\nu}{}_{\rho\sigma}$ is the second-order kinetic term. However, solving the perturbed field equation in this form is cumbersome, partly because it requires acting with the Green function 
\begin{equation}
D^{\mu\nu}{}_{\rho\sigma} = -\frac{1}{2 \partial^{2}}\big(\delta^{(\mu}{}_{\rho}\delta^{\nu)}{}_{\sigma}-\frac{1}{2} \eta^{\mu \nu} \eta_{\rho \sigma}\big) , 
\end{equation}
which is the inverse of the kinetic operator, on each term. Instead, we add a trace of $\mathcal{G}^{\mu\nu}$ that eliminates the trace of $\mathfrak{h}$ in eq. \eqref{Gupup}. Then the kinetic term is given simply by $-\frac{1}{2} \Box \mathfrak{h}^{\mu\nu}$, and the corresponding Green function is also trivial.

Moreover, we lower one index of the field equation by contracting $\tilde{\mathfrak{g}}_{\mu\nu}$ so that one index is up and the other is down. This mixed-index form simplifies the perturbative equations relative to other index placements. For instance, it is well known that for metric perturbations of the usual Ricci tensor in Riemannian geometry, the perturbations of $R^{\mu}{}_{\nu} \equiv g^{\mu\rho}R_{\rho\nu}$, are simpler than $R_{\mu\nu}$ and $R^{\mu\nu} \equiv g^{\mu\rho}g^{\nu\sigma} R_{\rho\sigma}$. Combining these simplifications, we thus find it useful to introduce the following tensor $\mathcal{R}^{\mu}{}_{\nu}$ 
\begin{equation}
  \mathcal{R}^{\mu}{}_{\nu} \equiv \mathcal{G}^{\mu\rho} \tilde{\mathfrak{g}}_{\rho\nu} - \frac{1}{2} \delta^{\mu}{}_{\nu} \mathcal{G}^{\rho\sigma} \tilde{\mathfrak{g}}_{\rho\sigma}\,.
\label{}\end{equation}
One can check that to linear order in $\mathfrak{h}$ the expansion of $\mathcal{R}^{\mu}{}_{\nu}$ is simply given by
\begin{equation}
  \mathcal{R}^{\mu}{}_{\nu}|_{\mathcal{O}(\mathfrak{h}^{1})}
  = 
  -\frac{1}{2} \Box \mathfrak{h}^{\mu}{}_{\nu}\,,
\label{}\end{equation}
and the corresponding inverse of the kinetic operator is trivial. By eliminating the trace part of the above Einstein equations, the external source part on the right-hand side is also simplified to a single term
\begin{equation}
  \mathcal{R}^{\mu}{}_{\nu}
  =
  \frac{1}{\sqrt{-\tilde{\mathfrak{g}}}} \mathfrak{j}^{\mu\kappa} \tilde{\mathfrak{g}}_{\kappa\nu}\,.
\label{EoM_R}\end{equation}
We stress that this rewriting of the Einstein field equations leads to a set of equations that are in one-to-one correspondence with the original equations of motion. Since they are related
by multiplication with an invertible matrix we can go back and forth between the two. But as will be clarified below, our rewriting is particularly useful in the case of a localized energy-momentum source and when used in conjunction with dimensional regularization, where certain integrals become scale-free and therefore vanish.

The Einstein equations can be further simplified by employing a set of auxiliary fields. Each auxiliary field is a small block of fields, and it becomes part of a hierarchy. At the end, we will represent the pseudo energy-momentum tensor $\tau^{\mu\nu}$ in a two-layer structure. When the problem is solved in these two separate steps, each auxiliary field, once computed, can be recycled repeatedly in higher-order computations. This makes the recursive procedure more efficient.

The first layer defines the auxiliary fields as follows:
\begin{equation}
\begin{aligned}
  A^{\mu\nu\rho}
  &=
  \mathfrak{g}^{\mu\kappa} \partial_{\kappa} \mathfrak{g}^{\nu\rho}\,,
  \qquad
  \overline{A}^{\mu\nu\rho}
  =
  \big(\mathfrak{g}^{\mu\kappa} - \eta^{\mu\nu}\big) \partial_{\kappa} \mathfrak{g}^{\nu\rho}\,,
  \\
  d_{\mu\nu}
  &=
  d_{\mu}d_{\nu}\,,
  \qquad
  d_{\mu} 
  = -\frac{1}{2(D-2)} \tilde{\mathfrak{g}}_{\rho \sigma} \partial_{\mu} \mathfrak{g}^{\rho \sigma}\,.
\end{aligned}\label{}
\end{equation}
Equivalently, these can be represented in terms of the metric perturbations as
\begin{equation}
\begin{aligned}
  A^{\mu\nu\rho}
  &=
  - \eta^{\mu\kappa} \partial_{\kappa} \mathfrak{h}^{\nu\rho}
  + \mathfrak{h}^{\mu\kappa} \partial_{\kappa} \mathfrak{h}^{\nu\rho}\,,
  \qquad
  \overline{A}^{\mu\nu\rho}
  =
  \mathfrak{h}^{\mu\kappa} \partial_{\kappa} \mathfrak{h}^{\nu\rho}\,,
  \\
  d_{\mu\nu}
  &=
  d_{\mu}d_{\nu}\,,
  \qquad
  d_{\mu}
  =
  \frac{1}{2(D-2)} \Big(\eta_{\rho \sigma} \partial_{\mu} \mathfrak{h}^{\rho \sigma}
  + \tilde{\mathfrak{h}}_{\rho \sigma} \partial_{\mu} \mathfrak{h}^{\rho \sigma}\Big)\,,
\end{aligned}\label{}
\end{equation}
In the next layer, the pseudo energy-momentum tensor is represented in harmonic gauge as
\begin{equation}
\begin{aligned}
  \mathcal{\tau}^{\mu}{}_{\nu}
  &= 
  W^{\mu}{}_{\nu}
  - \frac{1}{2} Z^{\mu}{}_{\nu}
  + \frac{1}{4} \delta^{\mu}{}_{\nu} Z^{\kappa}{}_{\kappa}
  + 2(D-2) \mathcal{D}^{\mu}{}_{\nu}
  - 2\delta^{\mu}{}_{\nu} \mathcal{D}^{\kappa}{}_{\kappa}\,,
\end{aligned}\label{Einstein_eq}
\end{equation}
where
\begin{equation}
\begin{aligned}
  W^{\mu}{}_{\nu} 
  &=
  2\partial_{\kappa}\overline{A}^{[\kappa\mu]}{}_{\nu}
  +2 \partial_{\kappa}  \big(A^{[\kappa\mu]\lambda}\tilde{\mathfrak{h}}_{\lambda\nu}\big)\,,
  \\
  Z^{\lambda}{}_{\kappa}
  &=
  \big(2A^{\rho\lambda\sigma}- A^{\lambda\rho\sigma}\big)\partial_{\kappa}\tilde{\mathfrak{g}}_{\rho\sigma}\,,
  \\
  \mathcal{D}^{\lambda}{}_{\kappa} 
  &=
  \mathfrak{g}^{\lambda\rho} d_{\rho\kappa}\,.
\end{aligned}\label{}
\end{equation}

For $D=4$ dimensions, we can rewrite the equation as
\begin{equation}
  \tau^{\mu}{}_{\nu}
  =
    \frac{1}{2} W^{\mu}{}_{\nu}
  - \frac{1}{8} T^{\mu\kappa}_{\nu\lambda} Z^{\lambda}{}_{\kappa} 
  + T^{\mu\kappa}_{\nu\lambda} \mathcal{D}^{\lambda}{}_{\kappa} 
\label{}\end{equation}
where
\begin{equation}
  T^{\mu\kappa}_{\nu\lambda} = 2 \delta^{\mu}{}_{\lambda} \delta^{\kappa}{}_{\nu} - \delta^{\mu}{}_{\nu} \delta^{\kappa}{}_{\lambda}\,.
\label{T-tensor}\end{equation}
%

\subsection{The double series expansion}

In Section 2 we introduced the double series expansion of the Kerr metric in harmonic coordinates. As we discussed, 
an expansion only in $G$ is not convenient for explicit computations
(at least so if we employ our recursive formalism) because of the analytic complexity of needed Fourier transforms. Thus we further expand in the spin parameter $a$, working with a double series expansion in $G$ and $a$.

In the same way, we now perform the double expansion for all field variables in the equations of motion and their field products, organizing each quantity as a simultaneous expansion in $G$ and $a$. We thus write the double expansions of the metric perturbations in the form
\begin{equation}
\begin{aligned}
  \mathfrak{h}^{\mu\nu}
  =
  \sum_{m=1}^{\infty} \sum_{n=0}^{\infty} \mathfrak{h}^{\mu\nu}\big|_{m,n} G^{m}a^{n}\,,
  \qquad
  \tilde{\mathfrak{h}}_{\mu\nu} 
  =
  \sum_{m=1}^{\infty} \sum_{n=0}^{\infty} \tilde{\mathfrak{h}}_{\mu\nu}\big|_{m,n} G^{m}a^{n}\,.
\end{aligned}\label{h_expansion}
\end{equation}
For concise expressions that automatically keep track of the order of expansion, we find it useful to introduce a new product ``$\ast$'' between two coefficients of the double expansion, defined as follows:
\begin{equation}
  \big(\phi_{1}*\phi_{2}\big)\big|_{m,n} ~\equiv~ \sum_{p=1}^{m-1} \sum_{q=0}^{n} \phi_{1}\big|_{m-p,n-q} \phi_{2}\big|_{p,q} \,.
\label{}\end{equation}

Using this notation, we can represent the inverse relation constraint \eqref{metric_perturbation} in terms of the coefficients of the expansion
\begin{equation}
  \tilde{\mathfrak{h}}_{\mu\nu}\big|_{m,n}
  =
    \eta_{\mu\rho} \eta_{\nu\sigma} \mathfrak{h}^{\rho\sigma}\big|_{m,n}
  + \big(\tilde{\mathfrak{h}}_{\mu\rho} * \mathfrak{h}^{\rho\sigma}\big)\big|_{m,n} \eta_{\sigma\nu}\,.
\label{inverse_metric_perturb}\end{equation}
From this relation we can derive the $\tilde{\mathfrak{h}}_{\mu\nu}\big|_{m,n}$ recursively. We substitute these results into the perturbative Einstein equation and iterate.

We therefore expand the Einstein field equation \eqref{Einstein_eq} using the same formalism: 
\begin{equation}
  \mathcal{R}^{\mu}{}_{\nu}\big|_{m,n}
  =
  \big(g_{\nu\kappa}*\mathfrak{j}^{\mu\kappa} \big)\big|_{m,n}
\label{}\end{equation}
where $g_{\mu\nu}$ is the ordinary metric tensor 
\begin{equation}
  g_{\mu\nu}
  =\frac{1}{\sqrt{-\tilde{\mathfrak{g}}}} \tilde{\mathfrak{g}}_{\mu\nu}\,,
\label{}\end{equation}
and the Einstein tensor is also expanded,
\begin{equation}
\begin{aligned}
  \mathcal{R}^{\mu}{}_{\nu}
  &=
  \sum_{m=1}^{\infty} \sum_{n=0}^{\infty} G^{m} a^{n} \mathcal{R}^{\mu}{}_{\nu} \big|_{m,n}\,.
\end{aligned}\label{}
\end{equation}
It is convenient to divide $\mathcal{R}^{\mu}{}_{\nu}$ into a linear term and the pseudo energy-momentum tensor $\tau^{\mu}{}_{\nu}$, which represents the nonlinear self--interaction terms,
\begin{equation}
    \mathcal{R}^{\mu}{}_{\nu} = -\frac{1}{2} \Box \mathfrak{h}^{\mu}{}_{\nu}
  + \frac{1}{2} \tau^{\mu}{}_{\nu}\,.
\label{}\end{equation}
Ordering this equation as a tower of increasing powers of our two expansion parameters, the associated $(m,n)$--th order term is given by
\begin{equation}
  \mathcal{R}^{\mu}{}_{\nu}\big|_{m,n}
  = 
  - \frac{1}{2} \Box \mathfrak{h}^{\mu}{}_{\nu}\big|_{m,n}
  + \frac{1}{2} \tau^{\mu}{}_{\nu}\big|_{m,n}\,, 
\label{}\end{equation}
where we have raised the lower index using the flat Minkowski (inverse) metric $\eta^{\mu\nu}$. Here $\tau^{\mu}{}_{\nu}$ can be written as
\begin{equation}
\begin{aligned}
  \tau^{\mu}{}_{\nu}\big|_{m,n}
  &=
    W^{\mu}{}_{\nu}\big|_{m,n}
  - \frac{1}{4} T^{\mu\kappa}_{\nu\lambda} Z^{\lambda}{}_{\kappa}\big|_{m,n}
  + 2T^{\mu\kappa}_{\nu\lambda} D^{\lambda}{}_{\kappa}\big|_{m,n}\,,
\end{aligned}\label{}
\end{equation}
where
\begin{equation}
\begin{aligned}
  W^{\mu}{}_{\nu}\big|_{m,n}
  &=
  2\partial_{\kappa}\overline{A}^{[\kappa\mu]}{}_{\nu}\big|_{m,n} 
  + 2\partial_{\kappa} \big( A^{[\kappa\mu]\lambda}*\tilde{\mathfrak{h}}_{\lambda\nu} \big)\big|_{m,n}\,,
  \\
  Z^{\mu}{}_{\nu}\big|_{m,n}
  &=
  \big((2A^{\kappa\mu\lambda} -  A^{\mu\kappa\lambda})*\partial_{\nu}\tilde{\mathfrak{h}}_{\kappa\lambda}\big) \big|_{m,n} \,,
  \\
  D^{\mu}{}_{\nu}\big|_{m,n}
  &= 
    \eta^{\mu\kappa}d_{\kappa\nu}\big|_{m,n}
  - \big(\mathfrak{h}^{\mu\kappa} *d_{\kappa\nu}\big) \big|_{m,n}\,.
\end{aligned}\label{EoMHierarchy2}
\end{equation}
and
\begin{equation}
\begin{aligned}
  A^{\mu\nu\rho}\big|_{m,n}
  &=
  - \eta^{\mu\kappa}\partial_{\kappa} \mathfrak{h}^{\nu\rho}\big|_{m,n}
  + \big(\mathfrak{h}^{\mu\kappa} * \partial_{\kappa} \mathfrak{h}^{\nu\rho}\big)\big|_{m,n}\,, 
  \qquad
  \overline{A}^{\mu\nu\rho}\big|_{m,n}
  =
  \big(\mathfrak{h}^{\mu\kappa}* \partial_{\kappa} \mathfrak{h}^{\nu\rho}\big)\big|_{m,n}\,, 
  \\
  d_{\mu}\big|_{m,n}
  &=
    \frac{1}{4} \eta_{\kappa\lambda}\partial_{\mu} \mathfrak{h}^{\kappa\lambda}\big|_{m,n}
  + \frac{1}{4} \big(\tilde{\mathfrak{h}}_{\kappa\lambda} * \partial_{\mu} \mathfrak{h}^{\kappa\lambda}\big)\big|_{m,n}\,,
  \qquad
  d_{\mu\nu}|_{m,n}
  =
  \big(d_{\mu} * d_{\nu}\big) \big|_{m,n}\,.
\end{aligned}\label{EoMHierarchy1}
\end{equation}
Then the equations of motion at $G^{m}$ and $a^{n}$ orders is given by
\begin{equation}
  \Box \mathfrak{h}^{\mu}{}_{\nu}\big|_{m,n}
  =
  \tau^{\mu}{}_{\nu}\big|_{m,n}
  -2 \big(g_{\nu\kappa}*\mathfrak{j}^{\mu\kappa} \big)\big|_{m,n}
\label{}\end{equation}
As we show in Appendix \ref{Appendix:B}, the equations of motion greatly simplify for a stationary and localized source. In particular, the source, being mulitplied by a single power of $G$ in the equations of motion, contributes only at first post-Minkowskian order, while at higher orders in $G$ the equations no longer depend directly on the source term
\begin{equation}
\begin{aligned}
  1{\rm PM}:\quad & \Box\mathfrak{h}^{\mu}{}_{\nu}\big|_{1,n} = -2\mathfrak{j}^{\mu\nu}\big|_{1,n}
  \\
  m{\rm PM}:\quad & \Box\mathfrak{h}^{\mu}{}_{\nu}\big|_{m,n} = \tau^{\mu}{}_{\nu}\big|_{m,n}\,, \qquad m>1\,.
\end{aligned}\label{perturbedEoM}
\end{equation}
We are thus ready to initiate the recursions.


\section{Recursion Relations}

Here we show construction of recursion relations based on the equations of motion derived in the previous section, and show how these relations can be used to reconstruct the Kerr metric order by order. The key observation is that the recursion is naturally organized in terms of a current, defined in momentum space from the Fourier transform of the metric perturbation. The recursions are evaluated using convolution integrals in momentum space. In order to perform this we introduce Fourier transforms of all fields in the equations of motion.  We begin with the graviton currents $\mathfrak{J}^{\mu\nu}|_{\boldsymbol{\ell}}^{m,n}$ and $\tilde{\mathfrak{J}}_{\mu\nu}|_{\boldsymbol{\ell}}^{m,n}$, which are defined as the Fourier transform of the metric perturbations 
\begin{equation}
\begin{aligned}
  \mathfrak{h}^{\mu\nu}(\boldsymbol{x})\big|_{m,n}
  =
  \int_{\ell} \, e^{i \boldsymbol{\ell}\cdot \boldsymbol{x}}\, \mathfrak{J}^{\mu\nu}\big|^{m,n}_{\boldsymbol\ell}\,,
  \qquad
  \tilde{\mathfrak{h}}_{\mu\nu}(\boldsymbol{x})\big|_{m,n}
  =
  \int_{\ell} e^{i \boldsymbol{\ell}\cdot \boldsymbol{x}}\, \tilde{\mathfrak{J}}_{\mu\nu}\big|^{m,n}_{\boldsymbol\ell}\,,
\label{def_Currents}\end{aligned}
\end{equation}
where 
\begin{equation}
  \int_{\ell} = \int \frac{\mathrm{d}^{d}\ell}{(2\pi)^{d}}\,,
  \qquad
  d=3-2\epsilon\,.
\label{}\end{equation}

Similarly, we introduce the currents for the auxiliary fields in the equations of motion \eqref{EoMHierarchy1} and \eqref{EoMHierarchy2}. We denote the fields and their Fourier transforms as follows:
\begin{equation}
\begin{aligned}
  A^{\mu\nu\rho}\big|_{m,n}
  &{=}
  \int_{\ell} e^{i \boldsymbol{\ell}\cdot \boldsymbol{x}} A^{\mu\nu\rho}\big|^{m,n}_{\ell}\,,
  \quad
  \overline{A}^{\mu\nu\rho}\big|_{m,n}
  {=}
  \int_{\ell}e^{i \boldsymbol{\ell}\cdot \boldsymbol{x}} \overline{A}^{\mu\nu\rho}\big|^{m,n}_{\boldsymbol\ell}\,,
  \quad
  d_{\mu\nu}\big|_{m,n}
  {=}
  \int_{\ell}  e^{i \boldsymbol{\ell}\cdot \boldsymbol{x}} d_{\mu\nu}\big|^{m,n}_{\boldsymbol\ell}\,,
  \\
  W^{\mu}{}_{\nu}\big|_{m,n}
  &{=}
  \int_{\ell} e^{i \boldsymbol{\ell}\cdot \boldsymbol{x}} W^{\mu}{}_{\nu}\big|^{m,n}_{\boldsymbol\ell}\,,
  \quad
  Z^{\mu}{}_{\nu}\big|_{m,n}
  {=}
  \int_{\ell} e^{i \boldsymbol{\ell}\cdot \boldsymbol{x}} Z^{\mu}{}_{\nu}\big|^{m,n}_{\boldsymbol\ell}\,,
  \quad
  \mathcal{D}^{\mu}{}_{\nu}\big|_{m,n}
  {=}
  \int_{\ell} e^{i \boldsymbol{\ell}\cdot \boldsymbol{x}} \mathcal{D}^{\mu}{}_{\nu}\big|^{m,n}_{\boldsymbol\ell}\,.
\end{aligned}\label{}
\end{equation}
Substituting the above Fourier transforms of the fields into the equations of motion one obtains a set of relations among the corresponding currents. These relations are the recursions which represent the higher-order currents in $G$ and $a$ in terms of lower-order currents. They can be solved iteratively order by order.

Obviously, the equations of motion consist of products of fields in position space. After the Fourier transforms, such products are mapped into convolutions. Denoting the Fourier transform of a field $\phi(x)$ by $\mathcal{F}[\phi]$, the product of two fields $\phi_{1}(x)$ and $\phi_{2}(x)$ is represented by the convolution
\begin{equation}
  \mathcal{F}[\phi_{1}\phi_{2}] = \mathcal{F}[\phi_{1}]* \mathcal{F}[\phi_{2}]\,.
\label{}\end{equation}
As we will see in the next section, solving the recursions iteratively, each field is represented by a product of Green functions, so that the resulting convolution integrals are
similar to loop integrals. In order to have a concise expression, it is convenient to introduce a bracket notation for the convolution:
\begin{equation}
  [\phi_{1}*\phi_{2}]^{m,n}_{\ell}
  = \sum_{p=1}^{m-1} \sum_{q=0}^{n} \int_{k} \phi_{1}\big|^{m-p,n-q}_{\ell-k} \phi_{2}\big|^{p,q}_{k}\,.
\label{}\end{equation}
In the presence of derivative operators such as $\partial_{\mu_{1}}\cdots \partial_{\mu_{i}}\phi_{1} * \partial_{\nu_{j}}\cdots \partial_{\nu_{j}} \phi_{2}\big|_{m,n}$, we may insert the momentum operator $\hat{k}^{\mu}$ into the corresponding entry as
\begin{equation}
\begin{aligned}
  &\big[\, \hat{k}^{\mu_{1}}\cdots \hat{k}^{\mu_{i}} \phi_{1}* \hat{k}^{\nu_{1}}\cdots \hat{k}^{\nu_{j}} \phi_{2} \,\big]^{m,n}_{\ell}
  \\
  &= \sum_{p=1}^{m-1} \sum_{q=0}^{n} \int_{k} \big(\ell^{\mu_{1}}-k^{\mu_{1}}\big) \cdots \big(\ell^{\mu_{i}}-k^{\mu_{i}}\big) k^{\nu_{1}}\cdots k^{\nu_{j}} \phi_{1}\big|^{m-p,n-q}_{\ell-k} \phi_{2}\big|^{p,q}_{k}\,.
\end{aligned}\label{}
\end{equation}
%


We are now ready to derive the recursion relations in momentum space explicitly. We begin by solving the recursion corresponding to the constraint for the inverse relation \eqref{Inverse_relation_constraint}. Substituting the Fourier transform \eqref{def_Currents} into the perturbed inverse constraint \eqref{inverse_metric_perturb}, we obtain
\begin{equation}
  \tilde{\mathfrak{J}}^{\mu\nu}\big|^{m,n}_{\ell}
  =
  \mathfrak{J}^{\mu\nu}\big|^{m,n}_{\ell}
  + \big[\, \mathfrak{J}^{\mu\kappa} * \tilde{\mathfrak{J}}_{\kappa}{}^{\nu}\big]^{m,n}_{\ell}\,.
\label{inverse_current_constraint}\end{equation}
We can thus determine $\tilde{\mathfrak{J}}^{\mu\nu}|_{m,n|\ell}$ at each order iteratively and starting with
the first layer of the hierarchy of equations of motion. First we introduce the Fourier transforms of the auxiliary fields \eqref{EoMHierarchy2},
\begin{equation}
\begin{aligned}
  A^{\mu\nu\rho}\big|^{m,n}_{\ell}
  &=
  - i \ell^{\mu} \mathfrak{J}^{\nu\rho}\big|^{m,n}_{\ell}
  + i \big[\, \mathfrak{J}^{\mu\kappa}* \hat{k}_{\kappa}\mathfrak{J}^{\nu\rho}\,\big]^{m,n}_{\ell}\,,
  \quad
  \overline{A}^{\mu\nu\rho}_{(m,n)|\ell}
  =
  i \big[\, \mathfrak{J}^{\mu\kappa}* \hat{k}_{\kappa}\mathfrak{J}^{\nu\rho}\,\big]^{m,n}_{\ell}\,,
  \\
  d_{\mu}\big|_{(m,n)|\ell}
  &=
    \frac{i}{4} \ell_{\mu} \mathfrak{J}^{\kappa}{}_{\kappa}\big|^{m,n}_{\ell}
  + \frac{i}{4} \big[\, \tilde{\mathfrak{J}}_{\kappa\lambda}* \hat{k}_{\mu}\mathfrak{J}^{\kappa\lambda}\, \big]^{m,n}_{\ell}\,,
  \quad
  d_{\mu\nu}\big|^{m,n}_{\ell}
  =
  \big[\,d_{\mu} * d_{\nu}\,\big]^{m,n}_{\ell} \,.
\end{aligned}\label{}
\end{equation}
Next, the recursion for the second layer is given by
\begin{equation}
\begin{aligned}
  W^{\mu}{}_{\nu}\big|^{m,n}_{\ell} 
   &=
    2i\ell_{\kappa} \overline{A}^{[\kappa\mu]}{}_{\nu}\big|^{m,n}_{\ell}
  + 2i\ell_{\kappa} \big[\, A^{[\kappa\mu]\lambda} * \tilde{\mathfrak{J}}_{\lambda\nu}\,\big]^{m,n}_{\ell} \,,
  \\
  Z^{\mu}{}_{\nu}\big|^{m,n}_{\ell}
  &=
  i \big[(2A^{\kappa\mu\lambda}- A^{\mu\kappa\lambda})*\hat{k}_{\nu} \tilde{\mathfrak{J}}_{\kappa\lambda}\,\big]^{m,n}_{\ell}\,,
  \\
  D^{\mu}{}_{\nu}\big|^{m,n}_{\ell}
  &= 
    \eta^{\mu\kappa}d_{\kappa\nu}\big|^{m,n}_{\ell}
  - \big[\, \mathfrak{J}^{\mu\kappa} * d_{\kappa\nu}\big]^{m,n}_{\ell}\,.
\end{aligned}\label{}
\end{equation}
Finally, we derive the recursion relation for the graviton current $\mathfrak{J}^{\mu\nu} |_{m,n|\ell}$  from the perturbed equations of motion in \eqref{perturbedEoM} by Fourier transform
\begin{equation}
\begin{aligned}
  \mathfrak{J}^{\mu}{}_{\nu}\big|^{1,n}_{\ell} 
  &= 
  \frac{2}{\ell^{2}} \,\mathfrak{j}^{00}\big|^{1,n}_{\ell}\,,
  \\
  \mathfrak{J}^{\mu}{}_{\nu}\big|^{m,n}_{\ell}
  &=
  - \frac{1}{\ell^{2}} \tau^{\mu}{}_{\nu}\big|^{m,n}_{\ell}\,,\qquad m\geq2
\end{aligned}\label{graviton_current_recursion}
\end{equation}
where
\begin{equation}
  \tau^{\mu}{}_{\nu}\big|^{m,n}_{\ell}
  =
  W^{\mu}{}_{\nu}\big|^{m,n}_{\ell}
  - \frac{1}{4} T^{\mu\lambda}_{\nu\kappa} Z^{\kappa}{}_{\lambda}\big|^{m,n}_{\ell}
  + 2T^{\mu\kappa}_{\nu\lambda} D^{\kappa}{}_{\lambda}\big|^{m,n}_{\ell}
  \,.
\label{}\end{equation}

It is immediately clear that there is no explicit source contribution beyond order $G$. Since the right hand side of \eqref{EoM_R} contains a source term coupled with the metric, it should generate terms at higher orders in $G$. As a result, all corrections beyond leading order are effectively governed only by the pseudo-energy-momentum tensor, with no further explicit dependence on the external source.  


\subsection{Ambiguities in dimensional regularization}\label{Sec:4.2}
It is important to note that Fourier integrals can become singular in certain cases and must therefore be properly regularized. Here we employ dimensional regularization and analytically continue the spatial dimension to $d=3-2\epsilon$. According to the Fourier transform formulae in Appendix \ref{Sec:A.1}, the transform of an expression of the form $P(x,y,z)/r^{n}$, where $P(x,y,z)$ is an arbitrary polynomial in $x,y,z$, becomes singular in three dimensions for $n=3$ due to $\Gamma(0)$ factors. In the case of the Kerr metric, such singular Fourier transforms arise at 3PM order and higher. To avoid these spurious singularities, we therefore perform dimensional regularization to $d=3-2\epsilon$. After completing the calculation in momentum space, we return to position space and finally take the limit $\epsilon \to 0$. This is completely standard.

However, dimensional regularization raises a different and more subtle issue when used in Fourier transforms. In $d=3-2\epsilon$ dimensions, the naive three-dimensional identity $x^{2}+y^{2}+z^{2}=r^{2}$ no longer holds, introducing an ambiguity in the Fourier transform. To understand this issue let us consider two functions $f_{1} = (x^{2}+y^{2}-2z^{2})/r^{m}$ and $f_{2}=(r^{2}-3z^{2})/r^{m}$. Obviously, these functions are identical in three dimensions, but they can differ in $d=3-2\epsilon$ dimensions after Fourier transforms, according to the choice of $m$, and even after taking the limit $\epsilon \to 0$. Denoting these transforms by $\mathcal{F}(f_{1})$ and $\mathcal{F}(f_{2})$ respectively, we find that they are given by the following expressions
\begin{equation}
  \mathcal{F}(f_{1})
  =
  -\frac{\bigl(p_{1}^{2}+p_{2}^{2}-2p_{3}^{2}\bigr)\pi^{\frac{3}{2}-\epsilon}2^{-m-2\epsilon+5}\Gamma(\frac{7-m}{2}-\epsilon)}{\Gamma(\frac{m}{2})}
  |\boldsymbol{ \ell}|^{m+2\epsilon-7}\,,
\end{equation}
and
\begin{equation}
\begin{aligned}
  \mathcal{F}(f_{2})
  & =
  -\frac{\pi^{\frac{3}{2}-\epsilon}2^{5-m-2\epsilon}}{\Gamma(\frac{m}{2})}
  \bigg[
  \bigl(|\boldsymbol{\ell}|^{2}-3\ell_{3}^{2}\bigr)\Gamma(\tfrac{7-m}{2}-\epsilon)|\boldsymbol{\ell}|^{m+2\epsilon-7}
  + \epsilon\Gamma(\tfrac{5-m}{2}-\epsilon) |\boldsymbol{\ell}|^{m+2\epsilon-5}
  \bigg]\,.
\end{aligned}\label{tst2Fourier}
\end{equation}
As one can check, the last term in the above expression \eqref{tst2Fourier} precisely accounts for the difference between $\mathcal{F}(f_{1})$ and $\mathcal{F}(f_{2})$ in the limit $\epsilon \to 0$. We denote this term by 
\begin{equation}
  \Delta \mathcal{F}_{12}
  =
  \mathcal{F}(f_{1}) - \mathcal{F}(f_{2})
  =
  -\frac{\pi^{\frac{3}{2}-\epsilon}2^{5-m-2\epsilon}}{\Gamma(\frac{m}{2})}\epsilon\Gamma(\tfrac{5-m}{2}-\epsilon) |\boldsymbol{\ell}|^{m+2\epsilon-5}\,.
\label{discrepancy}\end{equation}
Since $\Delta F_{12}$ appears to be proportional to $\epsilon$, it would normally vanish in the limit of $\epsilon \to 0$. However, a more careful inspection reveals that the factor of $\Gamma(\frac{5-m}{2}-\epsilon)$ develops a simple pole when $m$ is an odd integer greater than $5$, $m=5,7, 9, 11, \cdots$. Taking this singular behavior into account, the overall product of $\epsilon$ with the gamma function will yield finite contributions in these cases 
\begin{equation}
  \Delta \mathcal{F}_{12}
  =
  (-1)^{\alpha+1}\frac{2^3 \pi \alpha}{(2\alpha+1)!} |\boldsymbol\ell|^{2\alpha-2} + \mathcal{O}(\epsilon^{1})\,, 
  \qquad
  \alpha\in \mathbb{Z}_{+}\,,
\label{}\end{equation}
where $m = 2\alpha + 3 $. This term thus does not vanish in the limit of $\epsilon \to 0$, and the two Fourier transforms remain inequivalent even though the original expressions coincide in three dimensions. At first sight, this mismatch between Fourier transforms may seem disturbing. Remarkably, however, the inverse Fourier transform of the discrepancy $\Delta \mathcal{F}_{12}$ is 
\begin{equation}
  \mathcal{F}^{-1}\big[\Delta \mathcal{F}_{12}\big] (x)
  =
  \frac{2 \epsilon }{(m-2) r^{m-2}}\,,
\label{}\end{equation}
and this vanishes as $\epsilon \to 0 $. 

In summary, even if $f_{1}\vert_{\epsilon\to 0}=f_{2}\vert_{\epsilon\to 0}$ in position space, their Fourier transforms need not coincide, $\mathcal{F}[f_{1}] \neq \mathcal{F}[f_{2}]$. Nevertheless, the inverse Fourier transform remains well defined, and the two expressions become equivalent after transforming back to position space while keeping $\epsilon$ finite and only taking the limit $\epsilon \to 0$ in the end:
\begin{equation}
  \mathcal{F}^{-1}\big[\mathcal{F}[f_{1}]\big]_{\epsilon\to0}
  =
  \mathcal{F}^{-1}\big[\mathcal{F}[f_{2}]\big]_{\epsilon\to0}\,.
\label{inverseRelation}\end{equation}
Clearly, the same issue arises on the Fourier-transformed side. To illustrate, let us consider the following momentum-space integral:
\begin{equation}
\begin{aligned}
  I = \int_{k} \frac{|\boldsymbol{k}|^{2}-k_{1}^{2}-k_{2}^{2}-k_{3}^{2}}{|\boldsymbol{k}|^{m}|\boldsymbol{\ell}-\boldsymbol{k}|^{n}}\,.
\end{aligned}\label{}
\end{equation}
It vanishes at 3 dimensions, but in $3-2\epsilon$ dimension we find a non-trivial result according to the formula \eqref{Bubble_integral}
\begin{equation}
  I=
  -\frac{2^{2 \epsilon -3} \pi^{\epsilon -\frac{3}{2}} \epsilon\,  \Gamma\big(\frac{5-m}{2}-\epsilon\big) \Gamma \big(\frac{5-n}{2}-\epsilon\big) \Gamma \big(\frac{m+n-5}{2}+\epsilon \big)}{\Gamma \left(\frac{m}{2}\right) \Gamma \left(\frac{n}{2}\right) \Gamma \left(-\frac{m}{2}-2 \epsilon -\frac{n}{2}+5\right)} 
  |\boldsymbol{\ell}|^{5-m-n-2 \epsilon}\,.
\label{}\end{equation}
Generically, this integral vanishes as $\epsilon \to 0$, but for a specific choice of $m$ and $n$, it reduces to a finite value due to the identity $\epsilon \Gamma(-\alpha-\epsilon)=-\frac{(-1)^\alpha}{\alpha!}+\mathcal{O}\left(\epsilon ^1\right)$, where $\alpha$ is a positive integer. 

We therefore adopt the following prescription:
\begin{enumerate}
  \item Prior to taking any Fourier transform, all expressions are lifted to position space in $d=3-2\epsilon$ dimensions. We are free to choose either $r^{2}$ or $x^{2}+y^{2}+ z^{2}$ as the starting representation. The final result will be independent of this choice.
  
  \item We then perform the dimensionally regularized Fourier transform to $3-2\epsilon$ dimensional momentum space, and the resulting loop integrals are evaluated directly in $d=3-2\epsilon$ dimensions. Throughout this procedure, three dimensional identities such as $r^{2}= x^{2}+y^{2}+ z^{2}$ and $|\boldsymbol{\ell}|^{2}= \ell_{1}^{2}+\ell_{2}^{2} +\ell_{3}^{2}$ are not used at any intermediate stage.
  \end{enumerate}
  
We emphasize that already at step 1 there is no preferred prescription for whether to start with $r^{2}$ or $x^{2}+y^{2}+z^{2}$ in dimensional regularization. The essential requirement is not the particular choice at the starting point, but rather the consistent application of the rules described in step 2. If the computation is carried out entirely within that prescription at each intermediate stage, the final result will be correct.

\subsection{Convolution theorem}
We now show that the convolution theorem stated above continues to hold in dimensional regularization. As we have already seen, the double series expansion of the Kerr metric is generically of the form $P(r,x,y,z)/r^{n}$, where $P(r,x,y,z)$ is an arbitrary polynomial in $r,x,y,z$. It is therefore sufficient to analyze a representative class of functions of this type. To this end, we introduce the following two functions
\begin{equation}
  F(x)=\frac{x^{n_{1}}y^{n_{2}}z^{n_{3}}}{r^{n}}\,,
  \qquad
  G(x)=\frac{x^{m_{1}}y^{m_{2}}z^{m_{3}}}{r^{m}}\,.
\label{eq:conv-FG-def}\end{equation}
What we want to show is the convolution theorem
\begin{equation}
  \mathcal{F}[FG](\ell)
  =
  \int_{k}\, \mathcal{F}[F](\ell-k)\mathcal{F}[G](k)\,.
\label{eq:conv-theorem}
\end{equation}
It is convenient to introduce the basis functions
\begin{equation}
  \Phi_{\mathbf n,\lambda}
  =
  x^{n_{1}}y^{n_{2}}z^{n_{3}}\, r^{-\lambda}\,,
  \qquad
  \mathbf n = (n_{1},n_{2},n_{3})\,,
\label{eq:conv-basis}\end{equation}
then $F$ and $G$ can be rewritten simply as $F(x)=\Phi_{\mathbf n,\lambda_{1}}(x)$ and $G=\Phi_{\mathbf m,\lambda_{2}}(x)$. Using the Fourier transform formula in Appendix \ref{Sec:A.1}, we have
\begin{equation}
  \mathcal{F}[\Phi_{\mathbf n,\lambda}](\ell)
  =
  \mathcal{N}_{\lambda} \, i^{|\mathbf{n}|}\, \partial_{\ell}^{\mathbf n}
  \Big(
  |\boldsymbol\ell|^{n-d}\Big)\,,
  \qquad
  |\mathbf{n}|=n_{1}+n_{2}+n_{3}\,,
\label{eq:conv-basis-ft}\end{equation}
where
\begin{equation}
  \partial_{\ell}^{\mathbf n}
  =
  \partial_{\ell_{1}}^{n_{1}} \partial_{\ell_{2}}^{n_{2}} \partial_{\ell_{3}}^{n_{3}} \,,
  \qquad
  \mathcal{N}_{\lambda}
  =
  \frac{2^{d-\lambda}\pi^{d/2}\Gamma(\frac{d-\lambda}{2})}{\Gamma\!\left(\frac{\lambda}{2}\right)}\,.
\label{}\end{equation}

On the position-space side, the product of two basis elements is again a basis element,
\begin{equation}
  \Phi_{\mathbf n,n}(x)\,
  \Phi_{\mathbf m,m}(x)
  =
  \Phi_{\mathbf n+\mathbf m,n+m}(x),
\end{equation}
Hence the left-hand side of \eqref{eq:conv-theorem} is
\begin{equation}
  \mathcal{F}[FG](\ell)
  =
  \mathcal{F}[{\Phi}_{\mathbf n+\mathbf m,n+m}](\ell)
  =
  \mathcal{N}_{n+m}\,
  i^{|\mathbf n|+|\mathbf m|}\,
  \partial_{\ell}^{\mathbf n+\mathbf m}\Big(|\boldsymbol\ell|^{n+m-d}\Big)\,.
\label{eq:conv-lhs}
\end{equation}
Let us consider now the right hand side of \eqref{eq:conv-theorem}. Substituting the Fourier transform formula again \eqref{eq:conv-basis-ft}, we have
\begin{equation}
  \mathcal{N}_{n}\mathcal{N}_{m}
  i^{|\mathbf n|+|\mathbf m|}
  \int_{k}\,
  \partial_{\ell}^{\mathbf n}\Big(|\boldsymbol{\ell-k}|^{n-d}\Big)\,
  \partial_{k}^{\mathbf m}\Big(|\boldsymbol{k}|^{m-d}\Big)\,.
\label{eq:conv-rhs-start}\end{equation}
Because of the identity $\partial_{k_{i}}f(\ell-k)=-\partial_{\ell_{i}}f(\ell-k)$
we have
\begin{equation}
  \int d^{d}k\,\partial_{\ell}^{\mathbf n}f(\ell-k)\,\partial_{k}^{\mathbf m}g(k)
  =
  \partial_{\ell}^{\mathbf n+\mathbf m}
  \int d^{d}k\, f(\ell-k)g(k)\,.
\label{eq:conv-ibp-identity}\end{equation}
Applying this to \eqref{eq:conv-rhs-start} gives
\begin{equation}
  \int_{k} \mathcal{F}[F](\ell-k)\,\mathcal{F}[G](k)
  =
  \mathcal{N}_{n}\mathcal{N}_{m}\,i^{|\mathbf n|+|\mathbf m|}\partial_{\ell}^{\mathbf n+\mathbf m}
  J_{n,m}(\ell),
\label{eq:conv-rhs-reduced}\end{equation}
where
\begin{equation}
  J_{n,m}(\ell)
  =
  \int_{k}
  \frac{1}{|\boldsymbol{\ell-k}|^{d-n}|\boldsymbol{k}|^{d-m}}.
\label{eq:conv-J-def}
\end{equation}

Using the bubble integral formula in \eqref{Bubble_integral} for $n_{1}=n_{2}=n_{3}=0$ case, we obtain $J_{n,m}(\ell)$ 
\begin{equation}
  J_{n,m}(\ell)
  =
  \mathcal{C}_{n,m} |\boldsymbol\ell|^{n+m-d}\,,
  \qquad
  \mathcal{C}_{n,m}
  =
  \frac{1}{(4\pi)^{d/2}}
  \frac{
  \Gamma\!\left(\frac{d-n-m}{2}\right)
  \Gamma\!\left(\frac n2\right)
  \Gamma\!\left(\frac m2\right)
  }{
  \Gamma(\frac{d-n}{2})\Gamma(\frac{d-m}{2})
  \Gamma\!\left(\frac{n+m}{2}\right)
  }\,.
\label{eq:conv-J-scalar}\end{equation}
Interestingly, if we multiply all the factors, one can easily show
\begin{equation}
  \mathcal{N}_{n}\mathcal{N}_{m} \mathcal{C}_{n,m}
  =
  \mathcal{N}_{n+m}\,.
\label{eq:conv-gamma-identity}
\end{equation}
Combining the results, we find that the right hand side of eq. \eqref{eq:conv-theorem} equals
\begin{equation}
  \int_{k} \mathcal{F}[F](\ell-k)\,\mathcal{F}[G](k)
  =
  \mathcal{N}_{n+m}
  \,i^{|\mathbf n|+|\mathbf m|}
  \partial_{\ell}^{\mathbf n+\mathbf m}\Big(|\boldsymbol\ell|^{n+m-d}\Big)\,.
\label{eq:conv-rhs-final}
\end{equation}
This matches exactly\eqref{eq:conv-lhs}, thus demonstrating the convolution theorem \eqref{eq:conv-theorem}.

\section{Solving the Recursion Relations}

We now systematically compute the currents using the recursion relations with the prescription introduced in the previous section. We will show the solution up to fourth order in $G$ and to all orders in $a$. The recursion relations permit us to continue to higher and higher orders in the double expansion but since an all-order summation in $G$ at this stage seems
out of reach we see no point in proceeding further in the $G$-expansion, especially since we find a perfect match to this order with the compact expression expanded in $G$. The all-order sum of the Schwarzschild case \cite{Damgaard:2024fqj} gives high confidence that also the perturbatively computed Kerr metric has the expected radius of convergence in $1/r$.

Before solving the recursion relation, we discuss a subtle but important feature of the Fourier transform that is closely related to the gauge transformations that arise in solving the recursion and should be removed at the end. Suppose we solve the recursion under dimensional regularization and then perform the inverse Fourier transform to derive the corresponding metric. From 3PM order, if one solves the recursion and performs the inverse Fourier transform at $d = 3 - 2\epsilon$, the pseudo energy-momentum tensor is found to contain residual terms of the form $f^{\mu}{}_{\nu}(\delta r^2)$ that vanish in the limit $\epsilon \to 0$. Here $\delta r^{2}$ is the difference between $3-2\epsilon$ and 3-dimensional radial distance, $\delta r^{2} \equiv r^2 - x^2 - y^2 - z^2$. If we denote the pseudo energy-momentum tensor obtained by solving the recursion as $\hat{\tau}^{\mu}{}_{\nu}$, it splits into the  three-dimensional Einstein equation part $\tau^{\mu}{}_{\nu}$ and a remainder $f^{\mu\nu}(\delta r^2)$:
\begin{equation}
  \hat{\tau}^{\mu}{}_{\nu}
  =
    \tau^{\mu}{}_{\nu}
  + f^{\mu}{}_{\nu}(\delta r^2)\,.
\label{EoMWithResidue}\end{equation}
There is no problem up to this point since $f^{\mu}{}_{\nu}$ vanishes identically in three dimensions. However, to obtain the metric explicitly, we must act with the inverse Laplacian $\Box^{-1}$
\begin{equation}
  \mathfrak{h}^{\mu}{}_{\nu}
  =
    \Box^{-1} \tau^{\mu\nu}
  + \Box^{-1} f^{\mu\nu}(\delta r^2).
\label{GaugeTermSource}\end{equation}
Interestingly, in some cases $\Box^{-1} f^{\mu\nu}(\delta r^2)$ does not vanish in the $\epsilon \to 0$ limit, even though $f^{\mu\nu}(\delta r^2)$ itself vanishes in that limit. As an illustrative example, consider $\delta r^2/r^m$.
%
%
For that case the result remains finite as $\epsilon \to 0$ for $m = 5$ and $m=7$. This implies that even if $f^{\mu}{}_{\nu}(\delta r^2)$ vanishes in the limit $\epsilon \to 0$, the quantity $\Box^{-1} f^{\mu}{}_{\nu}(\delta r^2)$ can leave a finite remnant. In such cases $\Box^{-1}\hat{\tau}^{\mu}{}_{\nu}$ and $\Box^{-1}\tau^{\mu}{}_{\nu}$ will differ by finite terms in the limit
$\epsilon \to 0$. The difference will precisely correspond to a residual gauge transformation in harmonic gauge.
In this particular case, to obtain a metric free of such a gauge ambiguity, we first determine $\hat{\tau}^{\mu}{}_{\nu}$ before acting with $\Box^{-1}$, and then take the limit $\epsilon \to 0$.  This procedure systematically removes all contributions from $f^{\mu}{}_{\nu}$ and leaves only $\tau^{\mu}{}_{\nu}$. The graviton current and corresponding metric can then be written as
\begin{equation}
  \mathfrak{J}^{\mu}{}_{\nu}\big|^{n,m}_{\ell}
  =
  - |\boldsymbol\ell|^{-2}\, \tau^{\mu}{}_{\nu}\big|^{n,m}_{\ell}\,,
  \qquad
  \mathfrak{h}^{\mu}{}_{\nu}\big|_{n,m}
  =
  \mathcal{F}^{-1}\Big[- |\boldsymbol\ell|^{-2}\, \tau^{\mu}{}_{\nu}\big|^{n,m}_{\ell}\Big]\,.
\label{}\end{equation}
%


The terms to leading order in $G$ provide the initial values for solving to higher orders by means of recursion. In turn, the order-$G$ currents are completely determined by the external source derived above. Since only $\mathfrak{j}^{00}$ and $\mathfrak{j}^{0i}$ are non-vanishing in \eqref{00sourceMomentum}, $\mathfrak{j}^{ij}\big|^{1,m}_{\boldsymbol{\ell}}$ vanishes for all values of $m$. We begin by considering the case of $\mathfrak{J}^{00}\big|^{1,2n}_{\ell}$ where we find
\begin{equation}
\begin{aligned}
  \mathfrak{J}^{00}\big|^{1,2n}_{\ell}
  &=
  16\pi M \frac{(-1)^{n}}{(2n)!} \frac{(\ell_{\perp})^{2n}}{|\boldsymbol\ell|^{2}}
  =
  16\pi M \frac{(-1)^{n}}{(2n)!} \frac{\big(|\boldsymbol\ell|^{2}-\ell_{3}^{2}\big)^{n}}{|\boldsymbol\ell|^{2}}\,.
\end{aligned}\label{}
\end{equation}
By expanding $\big(|\boldsymbol{\ell}|^{2} - \ell_{3}^{2}\big)^{n}$ term, we can separate the expression into parts whose inverse Fourier transforms are trivial when reverting to three dimensions, and those that are not:
\begin{equation}
\begin{aligned}
  \mathfrak{J}^{00}\big|^{1,2n}_{\ell}
  &=
  \frac{16\pi M}{(2n)!} (-1)^{n}
  \bigg(
      (-1)^{n}\frac{(\ell_{3})^{2n}}{|\boldsymbol\ell|^{2}}
    + (-1)^{n-1} n \ell_{3}^{2 n-2} + \cdots
  \bigg)\,,
  \\
  &=
  \frac{16\pi M}{(2n)!} \bigg(\frac{(\ell_{3})^{2n}}{|\boldsymbol\ell|^{2}}+ \Delta^{00}\bigg)\,,
\end{aligned}\label{}
\end{equation}
where $\Delta^{00}$ denotes terms that all will vanish after taking the inverse Fourier transform and the limit $\epsilon \to 0$:
 $\mathcal{F}^{-1}\big[\Delta^{00}\big]_{\epsilon\to0} = 0$. It will simplify subsequent recursions to ignore such "null" terms, and we will do so in the following. In fact, this is only a first instance
 of ubiquitous scheme-dependent terms that arise in dimensional regularization. The appearance of such ambiguities is always directly linked to residual
 gauge invariance in harmonic gauge.
 
 It is interesting to compare the above result for $\hat{\mathfrak{J}}^{00}\big|^{1,2n}_{\ell}$ with what we would get if we instead use the all-order metric of eq. \eqref{1PM_h} as 
 truncated to order $G$, and take its Fourier transform in $d=3-2\epsilon$ dimensions. This yields
 \begin{equation}
 \mathfrak{J}^{00}\big|^{1,2n}_{\ell} =
  \frac{2^{4-2\epsilon}M\,\pi^{1-\epsilon}\, \Gamma(1-\epsilon)}{(2n)!} \frac{(\ell_{3})^{2n}}{|\boldsymbol{\ell}|^{2-2\epsilon}}
 \end{equation}
 which clearly differs from the expression we find above. However, the difference between the two expressions vanishes when taking the inverse Fourier transform and reverting to
 three dimensions. So the two expressions in momentum space are equivalent in this sense, and it is a matter of convenience which one is used.
 
 Similarly, for the components $\mathfrak{J}^{0i}\big|^{1,2n+1}_{\ell}$ we find that a convenient representation is given by
 \begin{equation}
 \mathfrak{J}^{0i}\big|^{1,2n+1}_{\ell} =
  \frac{i\, 2^{3-2\epsilon} M \pi^{1-\epsilon}}{(2n+1)!} \frac{\epsilon^{ij3}\ell_{j} (\ell_{3})^{2n}}{|\boldsymbol\ell|^{2-2\epsilon}}~.
 \end{equation}
 Both of these 1PM currents are readily summed to all orders in $a$:
 \begin{equation}
\begin{aligned}
  \mathfrak{J}^{00}\big|^{1}_{\ell}
  &=
  \sum_{n=0}^{\infty} a^{n}\,\mathfrak{J}^{00}\big|^{1,2n}_{\ell}
  = \frac{2^{4-2\epsilon}M\,\pi^{1-\epsilon}\, \Gamma(1-\epsilon)}{|\boldsymbol{\ell}|^{2-2\epsilon}} \cosh \big(a \ell_{3}\big)\,,
  \\
  \mathfrak{J}^{0i}\big|^{1}_{\ell}
  &=
  \sum_{n=0}^{\infty} a^{n}\,\mathfrak{J}^{0i}\big|^{1,2n}_{\ell}
  = \frac{i\, 2^{3-2\epsilon}M\,\pi^{1-\epsilon}\, \Gamma(1-\epsilon)}{|\boldsymbol{\ell}|^{2-2\epsilon}\ell_{3}} \epsilon^{ij3}\ell_{j} \sinh\big(a \ell_{3}\big)\,,
\end{aligned}\label{}
\end{equation}

We now proceed to solve the recursion relations by first determining the currents at order $G^2$. It is straightforward to solve these recursion relations by using the 1PM results shown above and evaluating the brackets using the bubble loop integral formula \eqref{Bubble_integral} given in Appendix A. Substituting the results into eq. \eqref{graviton_current_recursion}, we find $\mathfrak{J}^{\mu\nu}|^{2,n}_{\ell}$.

We here present the general expression $\mathfrak{J}^{00}|^{2,n}_{\ell}$ for arbitrary order in $a$,
\begin{equation}
\begin{aligned}
  \mathfrak{J}^{00}\big|^{2,2n}_{\ell}
  &=
  \frac{(-1)^{n}M^{2} \pi^{\frac{3}{2}-\epsilon} \Gamma\left(\frac{1}{2}-\epsilon-n\right)}{2^{3 n+2 \epsilon-1}(n!)^{2}|\boldsymbol{\ell}|^{1-2\epsilon}} 
  \\&\quad
  \times\sum_{k=0}^{n} (-1)^{k}\binom{n}{k}(6 k-7)(2 k-3)!!\,
  (\ell_{3})^{2n-2k} |\boldsymbol{\ell}|^{2k}\,
  \prod_{j=k}^{n-1}(2 \epsilon+2 j+1) \,.
\end{aligned}\label{}
\end{equation}
The $k$-sum can be evaluated explicitly to give the following compact form
\begin{equation}
\begin{aligned}
  \mathfrak{J}^{00}\big|^{2,2n}_{\ell}
  &=
  \frac{7 M^2\pi^{\frac{3}{2}-\epsilon}\,\Gamma (\frac{1}{2}-\epsilon)}{2^{2\epsilon+2n-1}(n!)^2}
  \frac{(\ell_3)^{2n}}{|\boldsymbol\ell|^{1-2\epsilon}}\
  {}_3F_2\!\left[\begin{matrix}-n,\,-\frac12,\,-\frac16\\-\frac76,\,\frac12+\epsilon\end{matrix};\frac{|\boldsymbol\ell|^{2}}{\ell_3^2}\right]\,,
\end{aligned}\label{}
\end{equation}
where $_3F_2$ and $_2F_1$ are hypergeometric functions. Expressions for the other components of the currents become quite lengthy, and we therefore summarize those
other components of the 2PM currents to arbitrary order in $a$ in Appendix \ref{Appendix:D.2}. We have explicitly checked the results up to $a^{16}$-order.

As discussed in Section \ref{sec:source} (and also in detail in ref. \cite{Damgaard:2026kqg}), the specific harmonic gauge metric of ref. \cite{Lin:2014laa} contains contributions that we cannot obtain by solving the recursion relations. We encounter these already at order $G^2$ for the components $ \mathfrak{J}^{11}_{(2,2n+1)|\ell}$ were, as indicated, they are given by all odd powers in $a$. We verify that they correspond to terms allowed by the residual gauge freedom but which are not generated by our recursion prescription. We can isolate these terms and extract their current form via the Fourier transform as follows:
\begin{equation}
\begin{aligned}
  \mathfrak{J}^{11}_{(2,2n+1)|\ell} 
  &=
  - \frac{M^2 2^{4-2\epsilon}\pi^{1-\epsilon} \Gamma(1-\epsilon)\,(n+1)}{(2n+3)!} \frac{\ell_1\ell_2\, \ell_3^{2n}}{\left|\boldsymbol{\ell}\right|^{2-2\epsilon}} \,,
  \\
  \mathfrak{J}^{22}_{(2,2n+1)|\ell}
  &=
  - \mathfrak{J}^{11}_{(2,2n+1)|\ell} \,,
  \\
  \mathfrak{J}^{33}_{(2,2n+1)|\ell}
  &= 0\,,
  \\
  \mathfrak{J}^{12}\big|^{2,2n+1}_{\ell}
  &=
  \frac{M^2 2^{3-2\epsilon} \pi^{1-\epsilon} \Gamma(1-\epsilon)\,(n+1)}{(2n+3)!}
  \frac{\left(\ell_1^2-\ell_2^2\right)\ell_3^{2n}}{\left|\boldsymbol{\ell}\right|^{2-2\epsilon}}\,,
  \\
  \mathfrak{J}^{i3}\big|^{2,2n+1}_{\ell}
  &=
  - \frac{M^2 2^{3-2\epsilon} \pi^{1-\epsilon}\Gamma(1-\epsilon)(n+1)}{(2n+3)!} 
  \frac{\epsilon^{ij3}\ell_j \ell_3^{2n+1}}{\left|\boldsymbol{\ell}\right|^{2-2\epsilon}}\,.
\end{aligned}\label{2PM_Jij_Odd_in_a}
\end{equation}
If we choose the gauge vector $\xi^{\mu}$ as
\begin{equation}
\begin{aligned}
  \xi^{0} &= 0\,,
  \qquad
  \xi^{1} = - \frac{M^2 2^{4-2\epsilon}\pi^{1-\epsilon} \Gamma(1-\epsilon)\,(n+1)}{(2n+3)!} \frac{\ell_2\, \ell_3^{2n}}{\left|\boldsymbol{\ell}\right|^{2-2\epsilon}}\,,
  \\
  \xi^{2} &= \frac{M^2 2^{4-2\epsilon}\pi^{1-\epsilon} \Gamma(1-\epsilon)\,(n+1)}{(2n+3)!} \frac{\ell_1\, \ell_3^{2n}}{\left|\boldsymbol{\ell}\right|^{2-2\epsilon}}
    \qquad
  \xi^{3} =0\,,
\end{aligned}\label{}
\end{equation}
we find that eq. \eqref{2PM_Jij_Odd_in_a} correspond to such trivial gauge terms that retain the metric in harmonic gauge but which are not generated with
our recursive prescription. This leads to the following interesting point:
When we solve the recursion relations to higher orders in $G$, we may either include or discard these pure-gauge terms. We observe that if we include them, the solution to the recursion relations agree with that of ref. \cite{Lin:2014laa} to the order we have checked. If they are excluded, the higher order results differ, and it is unclear if the result
of such an infinite-order calculation can be expressed in any simple closed form based on rational functions. Again, we stress that the difference is a choice of residual gauge 
symmetry only.

Let us now consider the currents at order $G^3$. As is clear, bubble integrals arise naturally when solving the recursion relations, and they exhibit divergences from this order, similar to what we encounter in the case of the Schwarzschild metric. Our needed Fourier transforms can also become divergent at this order. In fact, this is not a coincidence, as follows from the convolution theorem.  As discussed in the previous section, dimensional regularization is a consistent regularization scheme for this problem. 
Up to order $G^2$, all required bubble integrals are finite and no complications arise. Starting from order $G^3$, however, the ambiguities of dimensional regularization of Fourier transform and loop integrals discussed come into play, and we adhere to the prescription discussed there. In practice, our procedure is as follows. We first compute $\tau^{\mu}{}_{\nu}|^{3,n}_{\boldsymbol{\ell}}$, take its inverse Fourier transform and set $\epsilon \to 0$. This removes all terms proportional to $\delta r^{2} =r^2 - x^2 - y^2 - z^2$ and it chooses a particular fixing of the residual gauge freedom in harmonic coordinates.  To illustrate, consider collecting all terms depending on $\delta r^2$ at orders $a^0$ and $a^1$ at 3PM order,
\begin{equation}
\begin{aligned}
    f^{\mu}{}_{\nu}(\delta r^{2})\big|_{3,0}
  &=
  4M^3 \delta r^2 \begin{pmatrix}
    \frac{3 (11 r^2-2 \delta r^2)}{r^9} & 0 & 0 & 0 \\
    0&\frac{x^2-y^2-z^2}{r^9} & \frac{2x y}{r^9} & \frac{2x z}{r^9} \\
    0&\frac{2x y}{r^9}&\frac{y^{2}-x^2-z^2}{r^9} & \frac{2y z}{r^9} \\
    0&\frac{2x z}{r^9}&\frac{2y z}{r^9} & \frac{z^{2}-x^2-y^2}{r^9} \\
  \end{pmatrix}\,,
  \\
  f^{\mu}{}_{\nu}(\delta r^{2})\big|_{3,1}
  &=
  2 M^3 \delta r^2 \begin{pmatrix}
    0 & \frac{y (47 r^2-27 \delta r^2)}{r^{11}} & \frac{x (27 \delta r^2-47 r^2)}{r^{11}} & 0 \\
    \frac{y (52 r^2+27 \delta r^2)}{r^{11}} & 0 & 0 & 0 \\
    -\frac{x \left(27 \text{$\delta $r}^2+52 r^2\right)}{r^{11}} & 0 & 0 & 0 \\
    0 & 0 & 0 & 0 \\
  \end{pmatrix}\,.
\end{aligned}\label{}
\end{equation}
Acting with $\Box^{-1}$ on $\tau^{\mu}{}_{\nu}$ to obtain the desired $\mathfrak{h}^{\mu\nu}$, one finds that these terms do not vanish in general and correspond to residual gauge transformations
\begin{equation}
\begin{aligned}
  \Box^{-1}f^{\mu}{}_{\nu}\big|_{3,0}
  &=
  \left(\begin{array}{cccc}
   -\frac{276 M^3}{35 r^3} & 0 & 0 & 0 \\
   0 & -\frac{4 M^3 \left(r^2-6 x^2\right)}{105 r^5} & \frac{8 M^3 x y}{35 r^5} & \frac{8 M^3 x z}{35 r^5} \\
   0 & \frac{8 M^3 x y}{35 r^5} & -\frac{4 M^3 \left(r^2-6 y^2\right)}{105 r^5} & \frac{8 M^3 y z}{35 r^5} \\
   0 & \frac{8 M^3 x z}{35 r^5} & \frac{8 M^3 y z}{35 r^5} & -\frac{4 M^3 \left(r^2-6 z^2\right)}{105 r^5} \\
  \end{array}\right)\,,
  \\
  \Box^{-1}f^{\mu}{}_{\nu}\big|_{3,1}
  &=
  \left(\begin{array}{cccc}
   0 & -\frac{2 M^3 y}{r^5} & \frac{2 M^3 x}{r^5} & 0 \\
   -\frac{128 M^3 y}{35 r^5} & 0 & 0 & 0 \\
   \frac{128 M^3 x}{35 r^5} & 0 & 0 & 0 \\
   0 & 0 & 0 & 0 \\
  \end{array}\right)\,.
\end{aligned}\label{}
\end{equation}
We therefore eliminate all $\delta r^2$ terms following the prescription outlined at the beginning of this section, and then perform the Fourier transforms to obtain the currents. It is straightforward to obtain the general term for arbitrary powers of $a$:
\begin{equation}
\begin{aligned}
  \mathfrak{J}^{00}\big|^{3,2n}_{\ell}
  &=
  \frac{2^{4-2\epsilon} M^3 \pi^{1-\epsilon} \Gamma(-\epsilon)}{(2n+1)!}
  \frac{(\ell_3)^{2n}}{\left|\boldsymbol{\ell}\right|^{-2\epsilon}}\,,
  \\
  \mathfrak{J}^{0i}\big|^{3,2n+1}_{\ell}
  &=
  i\,\frac{2^{3-2\epsilon}M^3 \pi^{1-\epsilon} \Gamma(-\epsilon)(n+1)}{(2n+3)!}
  \frac{\epsilon^{ij3} \ell_j\,\ell_3^{2n}}{\left|\boldsymbol{\ell}\right|^{-2\epsilon}} \,.
\end{aligned}\label{}
\end{equation}
It is obvious that these currents all diverge as $\epsilon \to 0$. However, taking the inverse Fourier transform while keeping $\epsilon$ finite, they become regular and reproduce precisely the 3PM metric perturbations given in \eqref{Perturbation_h00} and \eqref{Perturbation_h0i}.


We finally turn to the order $G^4$. We derive the currents by solving the recursion relations, identify the pattern that emerges at each successive power of $a$, and read off the all-order general term, which we always verify explicitly up to order $a^{16}$.
From order $G^4$ the gauge terms at 2PM order for the $(ij)$-components (see eq. \eqref{2PMgauge}) begin to affect the solution of the recursion relations. There is evidently an infinity of possible ways to proceed, but we treat two cases explicitly here in order to illuminate the issue. First, we consider the case in which the 2PM gauge terms are included. In this case, solving the recursion yields currents consistent with the harmonic coordinate metric in Appendix \ref{Appendix:C}. One quickly identifies a pattern in the series expansion of the currents, and the general expression of $\mathfrak{J}^{00} \big|^{4,2n}_{\ell}$ in all orders  in $a$ is as follows:
\begin{equation}
\begin{aligned}
  \mathfrak{J}^{00} \big|^{4,2n}_{\ell}
  &=
  \frac{M^{4} \pi^{\frac{3}{2}-\epsilon} \Gamma\!\left(-\frac{1}{2}-\epsilon-n\right) }{2^{3 n+2 \epsilon-2} n!(n+1)!} |\boldsymbol{\ell}|^{1+2\epsilon} 
  \\&\quad\times
  \sum_{k=0}^{n}\binom{n}{k} (-1)^{n+k} (k-1)(2 k-3)!! (\ell_{3})^{2n-2k} |\boldsymbol{\ell}|^{2k} \prod_{j=k+1}^{n}(2 \epsilon+2 j+1)\,.
\end{aligned}\label{}
\end{equation}
As before, we can recast it in terms of hypergeometric functions
\begin{equation}
\begin{aligned}
  \mathfrak{J}^{00} \big|^{4,2n}_{\ell}
  &=
  \frac{M^{4} \pi^{\frac{3}{2}-\epsilon} \Gamma\!\left(-\frac{1}{2}-\epsilon\right)}{2^{2n+2 \epsilon-2}n!(n+1)!} |\boldsymbol{\ell}|^{1+2\epsilon} (\ell_{3})^{2 n} 
  \\& 
  \times\left[{ }_{2} F_{1}\bigg[\begin{array}{c}-n,-\frac{1}{2} \\ \epsilon+\frac{3}{2}\end{array} ; \tfrac{|\boldsymbol{\ell}|^{2}}{\ell_{3}^{2}}\bigg]
  -\frac{n |\boldsymbol{\ell}|^{2}}{(2 \epsilon+3) \ell_{3}^{2}}\,{}_{2} F_{1}\bigg[\begin{array}{c}1-n, \frac{1}{2} \\ \epsilon+\frac{5}{2}\end{array} ; \tfrac{|\boldsymbol{\ell}|^{2}}{\ell_{3}^{2}}\bigg]\right]\,.
\end{aligned}\label{}
\end{equation}
The reader can find the other components in Appendix \ref{Appendix:D.2}. These results exactly match the all-order metric in \eqref{Perturbation_h00}. Similarly, as shown in the appendix, we also find complete agreement with the $(0i)$ and $(ij)$-components of the metric and \eqref{Perturbation_h0i}. 

Next let us finally consider the option at which the 2PM gauge terms are absent. We denote the currents as $\mathfrak{J}'^{\mu\nu}|_{4,n}$, and correspondingly denote those metric components by $\mathfrak{h}'^{\mu\nu}$ by Again focusing on the $(00)$-components, we read obtain 
\begin{equation}
\begin{aligned}
  \mathfrak{J}'^{00} \big|^{4,2n}_{\ell}
  &=
  \frac{M^{4} \pi^{\frac{3}{2}-\epsilon} \Gamma\left(-\frac{3}{2}-\epsilon\right)}{2^{2n+2 \epsilon} n!(n+1)!} |\boldsymbol{\ell}|^{1+2\epsilon}
  \\&\quad
  \times \sum_{k=0}^{n} \frac{(-n)_{k}}{\left(\epsilon+\frac{5}{2}\right)_{k} k!}\left[\frac{8 n+17}{2(n+2)}\left(-\tfrac{1}{2}\right)_{k}+\frac{3}{4}\left(\tfrac{1}{2}\right)_{k}+\left(\tfrac{3}{2}\right)_{k}\right] (\ell_{3})^{2n-2k} |\boldsymbol{\ell}|^{2k}\,.
  \\
  &=
  \frac{M^{4} \pi^{\frac{3}{2}-\epsilon} \Gamma\!\left(-\frac{3}{2}-\epsilon\right)}{2^{2n+2 \epsilon}n!(n+1)!} |\boldsymbol{\ell}|^{1+2\epsilon} (\ell_{3})^{2n} 
  \Bigg[~
    \frac{8 n{+}17}{2n{+}4}\, {}_{2} F_{1}\bigg[\begin{array}{c}-n,-\frac{1}{2} \\ \epsilon+\frac{5}{2}\end{array} ; \tfrac{|\boldsymbol{\ell}|^{2}}{\ell_{3}^{2}}\bigg]
  \\&\qquad\qquad\qquad\qquad\qquad\qquad
  + \frac{3}{4}{ }_{2} F_{1}\bigg[\begin{array}{c}-n, \frac{1}{2} \\ \epsilon+\frac{5}{2}\end{array} ; \tfrac{|\boldsymbol{\ell}|^{2}}{\ell_{3}^{2}}\bigg]
  + {}_{2} F_{1}\bigg[\begin{array}{l}-n, \frac{3}{2} \\ \epsilon+\frac{5}{2}\end{array} ; \tfrac{|\boldsymbol{\ell}|^{2}}{\ell_{3}^{2}}\bigg]~\Bigg]\,.
\end{aligned}\label{}
\end{equation}
Taking inverse Fourier transformation, we obtain the general $a^{n}$-th order term denoted as $\mathfrak{h}'^{00}|_{4,n}$,
\begin{equation}
\begin{aligned}
  \mathfrak{h}'^{00}\big|_{4,n}
  &=
  \frac{M^{4}}{r^{2n+4}}\bigg[
  \frac{23 n+48}{3(n+2)} P_{n}^{(0,1)}(u)-\frac{59n+126}{15(n+2)} P_{n-1}^{(0,2)}(u)
  \\&\qquad\qquad\quad
  +\frac{2 n+3}{n+2} \sum_{j=2}^{n} \frac{2^{j-2} j!}{(2 j+3)!!} P_{n-j}^{(0, j+1)}(u)
  \bigg]\,,
\end{aligned}\label{}
\end{equation}
where $u=1- 2z^2/r^{2}$ and $P^{(\alpha,\beta)}_{n} (u)$ is the Jacobi polynomial. We summarize the other components obtained by this prescription in Appendix \ref{Appendix:D.2}.

\section{Conclusion}

Generalizing the recursive method of ref. \cite{Damgaard:2024fqj} to the two-parameter metric of a Kerr black hole, we have explored the extent to which perturbation theory can be summed to infinite orders in both coupling $G$ and reduced spin $a = J/M$. Working in Landau-Lifshitz variables and harmonic gauge, it becomes evident already at low orders that the residual gauge freedom of harmonic coordinates plays a much bigger role than in the case of the Schwarzshild black hole \cite{Damgaard:2024fqj}. This was already observed in greater generality when applying the same recursive method to a general multipole expansion \cite{Damgaard:2026kqg}. We have demonstrated that in order to achieve a summation of the perturbative expansion into the specific form of the harmonic-gauge Kerr metric of eq. \eqref{LandauLifshitzExpression} is a matter of additional gauge choice beyond the gauge fixing into harmonic coordinates. The same phenomenon is operative in the case of the Schwarzschild metric \cite{Fromholz:2013hka} but the residual gauge freedom is much smaller in that case, and appears perturbatively only beyond third order in $G$. Here, in the Kerr case, a gauge ambiguity shows up already at order $G^2$ in the $\mathfrak{h}^{ij}$-components. From this point onward we can choose several paths towards an all-order computation. First, we can proceed iteratively, and develop the perturbative expansion of the metric, ignoring the gauge artefact mismatch with eq. \eqref{LandauLifshitzExpression} for the $\mathfrak{h}^{ij}$-components already at order $G^2$. There is no obstacle towards doing so but we will then clearly not be able to derive the harmonic gauge metric of eq. \eqref{LandauLifshitzExpression} in summed form. Indeed, we observe empirically that the recursions in that case tend to not yield recognizable patterns, an indication that the perturbative expansion of the metric in that case probably will not sum up to any simple rational function, such as that of eq. \eqref{LandauLifshitzExpression}. We stress that the totally of all Kerr metrics in harmonic gauge are on exactly the same footing: they are sourced by the same disk and ring-like energy-momentum tensor, they have correct fall-off at Minkowskian infinity, and they will yield the same physical observables (such as scattering angles, etc.). Although we clearly cannot prove it, we anticipate that the totality of all these gauge-equivalent harmonic gauge metrics will share the same region of convergence of the perturbative series.

An approach that takes us closer to the particular Kerr metric solution of eq. \eqref{LandauLifshitzExpression} is to introduce by hand the gauge vector that brings $\mathfrak{h}^{ij}$ to the form of eq. \eqref{LandauLifshitzExpression} at order $G^2$. Iterating further with this new order $G^2$ input for $\mathfrak{h}^{ij}$ leads to a much simpler perturbative series. We defined the currents via the Fourier transform of the metric perturbations and derived a recursion relation for them from the equations of motion. This recursion was solved iteratively up to 4PM order and to all orders in $a$, both with and without the inclusion of gauge artefacts at order $G^2$.

A key part of the analysis was clarifying ambiguities that arise in dimensional regularization when applied to Fourier transforms and loop integrals. Under dimensional regularization to $d =3-2\epsilon$ dimensions, $r^2$ and $x^2 + y^2 + z^2$ are not equal. After taking the Fourier transform, this discrepancy becomes more apparent, and even in the limit $\epsilon \to 0$, the difference does not vanish in general. However, we showed that after taking the inverse Fourier transform, these finite discrepancies disappear in the $\epsilon \to 0$ limit, and that these ambiguities therefore do not affect the final results. We also identified the mechanism by which gauge transformations can arise when solving the recursions and incorporated a procedure into our prescription to remove them, thereby preventing spurious gauge artefacts from entering through the Fourier transform. Finally, we proved that the convolution theorem holds under dimensional regularization, ensuring that our recursion relation correctly generates the metric perturbations order by order.

Although our discussion has been carried out up to fourth order in $G$ (and for arbitrary powers of $a$), the same strategy can clearly be extended to higher orders in $G$. In Appendix~\ref{Appendix:C}, we derive general expressions for arbitrary powers of both $G$ and the spin parameter $a$ by expanding eq. \eqref{LandauLifshitzExpression}. These results may therefore be used to construct a solution to any desired order. As we have discussed in great detail, residual gauge freedom means that this will only be one out of a multitude of gauge equivalent Kerr metrics but it should still be possible to determine the radius of convergence of the series, and presumably find that it will agree with the expression given in Appendix C.

\vspace{0.4cm}
\noindent
{\bf Acknowledgements}

We thank Thibault Damour for discussions. The work of HL is supported by KIAS Individual Grant PG105701. The work of KL is supported by the National Research Foundation of Korea(NRF) grant funded by the Korea government(MSIT) RS-2025-24533223 and KIAS Individual Grant QP106001. The work of TR is supported by KIAS Individual Grant QP108201 via the Quantum Universe Center at KIAS.

\appendix
\newpage

\section{Useful formulae}

\subsection{Fourier Integrals}\label{Sec:A.1}
Our convention of the (inverse) Fourier transform is as follows:
\begin{equation}
  H(x)=\int \frac{d^{d}\ell}{(2\pi)^{d}}\, e^{i\ell\cdot x}\widetilde{H}(\ell)\,.
  \qquad
  \widetilde{H}(\ell)=\int d^{d}x\, e^{-i\ell\cdot x}H(x)\,,
\label{eq:convFT}
\end{equation}
where $H(x)$ is an arbitrary function in position space, and $\widetilde{H}$ is the corresponding Fourier dual. The explicit formula of the Fourier transform of the $1/r^{n}$ and its inverse transform are given by
\begin{equation}
\begin{aligned}
  \mathcal{F}\bigg[\frac{1}{r^{n}}\bigg]
  &=
  \frac{\pi^{\frac{d}{2}} 2^{d-n}  \Gamma\left[\frac{d-n}{2}\right]}{\Gamma \left[\frac{n}{2}\right]} \frac{1}{|\boldsymbol{\ell}|^{d-n}}\,,
  \\
  \mathcal{F}^{-1}\bigg[\frac{1}{|\boldsymbol\ell|^{n}} \bigg]
  &=
  \frac{1}{2^{n} \pi^{\frac{d}{2}}} \frac{\Gamma(\frac{d-n}{2})}{\Gamma(\frac{n}{2})} \frac{1}{r^{d- n}}\,.
\end{aligned}\label{Fourier1}
\end{equation}
Here $|\boldsymbol\ell|$ and $r$ in $d=3-2\epsilon$ case are defined by
\begin{equation}
\begin{aligned}
  r^{2} &= x^{2}+y^{2}+z^{2} + |\boldsymbol{x}_{\perp}|^{2}
  \\
  |\boldsymbol\ell|^{2} &= \ell_{1}^{2}+\ell_{2}^{2}+\ell_{3}^{2} + |\boldsymbol\ell_{\perp}|^{2}
\end{aligned}\label{}
\end{equation}
where $|\boldsymbol{x}_{\perp}|^{2}$ and $|\boldsymbol\ell_{\perp}|^{2}$ are formal expression of the contributions from the extra $-2\epsilon$ dimensional space. 

An important feature of this result is that, for certain values of $n$, the formula above must be applied carefully, since the Fourier transform is not an ordinary rational function and should instead be understood as a distribution. In particular, when $n=0$ or a negative even integer, $n=0,-2,-4,-6,\cdots$ the factor $\Gamma[n/2]$ in the denominator on the right-hand side of \eqref{Fourier1} diverges, so that the result may appear to vanish trivially at first sight. However, this naive conclusion is not correct. If we take $n=0$, we should  recover a delta-function, $\mathcal{F}[1] = (2\pi)^{d}\delta^{(d)}(\boldsymbol\ell)$ instead of $0$. The important thing that should not be overlooked here is that $\mathcal{F}[r^{-n}]$ is also singular at $|\boldsymbol{\ell}|=0$ for those cases, and therefore cannot be set to zero as an ordinary function over the entire momentum space. Rather, it should be interpreted as a distribution: it vanishes away from $|\boldsymbol{\ell}|=0$, while retaining a nontrivial singular contribution localized at that point. 

More explicitly, let us consider the limit $n\to 0$ in $\mathcal{F}[r^{-n}]$. Using the expansion formula of the Gamma function, $\Gamma\left(\frac{n}{2}\right)=\frac{2}{n}-\gamma_{E}+\mathcal{O}(n)$, we have 
\begin{equation}
  \mathcal{F}[r^{n}]_{n\to 0} = 2^{d-1} \pi^{\frac{d}{2}} \Gamma\left(\tfrac{d}{2}\right) n|\boldsymbol{\ell}|^{n-d} +\mathcal{O}\left(n^{2}\right) \,.
\label{Fourier_r^0}\end{equation}
As we discussed before, this vanishes except at  $|\boldsymbol{\ell}| = 0$. In order to check that it is indeed a delta function, we multiply a test function $\phi(\ell)$ and integrate with respect to $\ell$ in a sphere with radius $R$
\begin{equation}
\begin{aligned}
  \int_{|\ell|<R} \mathrm{d}^{d} \ell\left(n|\ell|^{-d+n}\right) \phi(\ell)\,.
\end{aligned}\label{}
\end{equation}
Since $\mathrm{d}^{d}\ell = |\boldsymbol{\ell}|^{d-1} \mathrm{d} |\boldsymbol{\ell}| \mathrm{d} \Omega_{d-1}$ and $\phi(\boldsymbol{\ell}) \approx \phi(0)$ in the $n\to 0$ limit, the integral reduces to
\begin{equation}
\begin{aligned}
  &\phi(0) \Omega_{d-1} \int_{0}^{R} d|\boldsymbol\ell|\,|\boldsymbol\ell|^{d-1} n|\ell|^{-d+n}
  =
  \phi(0) \Omega_{d-1} R^{n}
  = \phi(0) \Omega_{d-1} 
\end{aligned}\label{}
\end{equation}
Using $\Omega_{d-1} = \frac{2 \pi^{d / 2}}{\Gamma(d / 2)}$, one can identify 
\begin{equation}
  \lim_{n\to0} n|\ell|^{-d+n} = \frac{2 \pi^{\frac{d}{2}}}{\Gamma(\frac{d}{2})} \delta^{(d)}(\boldsymbol\ell)\,,
\label{}\end{equation}
and we get $\mathcal{F}[1] = (2\pi)^{d} \delta(\boldsymbol{\ell})$ as required. In general, we can derive $\mathcal{F}\big[r^{2m}\big]$ and $\mathcal{F}^{-1}\big[r^{2m}\big]$ for $m = 0,1,2,3,\cdots$ by applying the Laplacian for $d$-dimensional momentum space
\begin{equation}
\begin{aligned}
  \mathcal{F}\big[r^{2m}\big]
  &=
  (2\pi)^{d} \left(-\Delta_{\ell}\right)^{m} \delta^{(d)}(\ell)\,,
  \qquad
  \Delta_{\ell} = \sum_{i=1}^{d} \partial^{2}_{\ell_{i}}\,,
  \\
  \mathcal{F}^{-1}\big[|\boldsymbol{\ell}|^{2m}\big]
  &=
  \left(-\Delta\right)^{m} \delta^{(d)}(\boldsymbol x)\,,
  \qquad
  \Delta = \sum_{i=1}^{d} \partial^{2}_{x_{i}}\,,
\end{aligned}\label{Fourier_spectial_cases}
\end{equation}
 and thus giving derivatives of delta-functions. 
We next consider the Fourier transform of the functions with nontrivial numerators. Since the numerators can be obtained by acting with derivative operators, we have the Fourier and inverse Fourier transform formula as
\begin{equation}
\begin{aligned}
  \mathcal{F}\bigg[\frac{x^{\alpha}y^{\beta}z^{\gamma}}{r^{n}}\bigg]
  &=
  i^{\alpha+\beta+\gamma} 
  \frac{\partial^{\alpha}}{\partial \ell_{1}^\alpha}
  \frac{\partial^{\beta}}{\partial \ell_{2}^\beta}
  \frac{\partial^{\gamma}}{\partial \ell_{3}^\gamma}
  \mathcal{F}\bigg[\frac{1}{r^{n}}\bigg]\,,
  \\ 
  \mathcal{F}^{-1}\bigg[\frac{\ell_{1}^{\alpha} \ell_{2}^{\beta} \ell_{3}^{\gamma}}{|\boldsymbol\ell|^{n}} \bigg]
  &=
  (-i)^{\alpha+\beta+\gamma} 
  \frac{\partial^{\alpha}}{\partial x^\alpha}
  \frac{\partial^{\beta}}{\partial y^\beta}
  \frac{\partial^{\gamma}}{\partial z^\gamma}
  \mathcal{F}^{-1}\bigg[\frac{1}{|\boldsymbol\ell|^{n}}\bigg]\,.
\end{aligned}\label{}
\end{equation}
Again, we need to be careful for $n = 0,-2,-4,-6\cdots$ cases, and it is given by in terms of \eqref{Fourier_spectial_cases}
\begin{equation}
\begin{aligned}
  \mathcal{F}\Big[x^{\alpha}y^{\beta}z^{\gamma}r^{2m}\Big]
  &=
  i^{\alpha+\beta+\gamma} 
  \frac{\partial^{\alpha}}{\partial \ell_{1}^\alpha}
  \frac{\partial^{\beta}}{\partial \ell_{2}^\beta}
  \frac{\partial^{\gamma}}{\partial \ell_{3}^\gamma}
  \mathcal{F}\big[r^{2m}\big] \,,
  \\ 
  \mathcal{F}^{-1}\bigg[\ell_{1}^{\alpha} \ell_{2}^{\beta} \ell_{3}^{\gamma}|\boldsymbol\ell|^{2m} \bigg]
  &=
  (-i)^{\alpha+\beta+\gamma} 
  \frac{\partial^{\alpha}}{\partial x^\alpha}
  \frac{\partial^{\beta}}{\partial y^\beta}
  \frac{\partial^{\gamma}}{\partial z^\gamma}
  \mathcal{F}^{-1}\big[|\boldsymbol{\ell}|^{2m}\big]\,.
\end{aligned}\label{}
\end{equation}
where $m=0,1,2,3,\cdots$.

\subsection{A bubble Integral Formula}

\begin{equation}
\begin{aligned}
  &I_{\lambda_{1},\lambda_{2}}[{n_{1},n_{2},n_{3}}]
  =
  \int \frac{\mathrm{d}^{d} k}{(2 \pi)^{d}} 
  \frac{(k_{1})^{n_{1}}(k_{2})^{n_{2}}(k_{3})^{n_{3}}}{\left|\boldsymbol{k}\right|^{2\lambda_{1}}
  	\left|\boldsymbol{\ell}-\boldsymbol{k}\right|^{2\lambda_{2}}} 
  \\
  &=
  \frac{1}{(4 \pi)^{d / 2} \Gamma(\lambda_{1}) \Gamma(\lambda_{2})}
  	\sum_{a_{1}=0}^{\lfloor n_{1} / 2\rfloor} 
  	\sum_{a_{2}=0}^{\lfloor n_{2} / 2\rfloor} 
  	\sum_{a_{3}=0}^{\lfloor n_{3} / 2\rfloor}
  	\binom{n_{1}}{2 a_{1}}
  	\binom{n_{2}}{2 a_{2}}
  	\binom{n_{3}}{2 a_{3}} \frac{(2 a_{1})!(2 a_{2})!(2 a_{3})!}{2^{2 r} a_{1}! a_{2}!a_{3}!} 
  \\&\quad
  \times
  \frac{
  	\Gamma\big(\lambda-\frac{d}{2}-r\big) 
  	\Gamma\big(\frac{d}{2}+r-\lambda_{2}\big) 
  	\Gamma\big(n-r+\frac{d}{2}-\lambda_{1}\big)}{\Gamma(n+d-\lambda)} 
  (\ell^{2})^{\frac{d}{2}-\lambda+r} 
  \ell_{1}^{n_{1}-2 a_{1}}
  \ell_{2}^{n_{2}-2 a_{2}}
  \ell_{3}^{n_{3}-2 a_{3}} 
\end{aligned}\label{Bubble_integral}
\end{equation}
where $n=n_{1}+n_{2}+n_{3}$, $\lambda = \lambda_{1}+\lambda_{2}$ and $r=a_{1}+a_{2}+a_{3}$.


\section{Derivation of \eqref{perturbedEoM}}\label{Appendix:B}

We now show that, even when the source term is dressed by additional fields as on the right-hand side of \eqref{EoM_R}, an arbitrary localized source contributes only at 1PM order under the stationary condition. A localized source $\mathfrak{j}^{\mu\nu}$ is expanded as
\begin{equation}
  \mathfrak{j}^{\mu\nu}(\vec{x})=\sum_{l=0}^{\infty} \frac{(-1)^{l}}{l!} M^{\mu\nu}_{L} \partial_{L} \delta^{(3)}(\vec{x})\,,
  \qquad
  M^{\mu\nu}_{L}=\int d^{3} x \, x^{L} \mathfrak{j}^{\mu\nu}(\vec{x})\,,
\label{}\end{equation}
where $L$ is a multi index with the length $l$, $L = i_{1} i_{2}\cdots i_{l}$ and $\partial_{L} = \partial_{i_{1}} \partial_{i_{2}} \cdots \partial_{i_{l}}$. In the Fourier basis, $\mathfrak{j}^{\mu\nu}(\vec{x})$ is represented as
\begin{equation}
  \mathfrak{j}^{\mu\nu}(\vec{x})
  =
  \sum_{l=0}^{\infty} \frac{(-i)^{l}}{l!} \int_{\vec{k}} e^{i\vec{k} \cdot \vec{x}} M^{\mu\nu}_{L} k_{L} \,.
\label{}\end{equation}
Here the crucial point is that the external source does not depend on the propagator. On the other hand, we consider an arbitrary scalar field $\phi(\vec{x})$ which has the following form:
\begin{equation}
  \phi(\vec{x})
  =
  \int_{\vec{k}} e^{i\vec{k}\cdot \vec{x}} \ \frac{P(\vec{k})}{|\vec{k}|^{n}}\,,
  \qquad
  n \in \mathbb{R}_{+}\,,
\label{conditionforField}\end{equation}
where $P(\vec{k})$ is an arbitrary polynomial of $k_{1}$, $k_{2}$ and $k_{3}$. Note that the fields appearing in the equations of motion must have this form for the stationary case due to the bubble integral structure. Then the product between $\mathfrak{j}^{\mu\nu}(\vec{x}) \phi(\vec{x})$ can be evaluated using the following convolution integral:
\begin{equation}
  \mathfrak{j}^{\mu\nu}(\vec{x}) \phi(\vec{x})
  =
  \sum_{l=0}^{\infty} \frac{(-i)^{l}}{l!}
  \int_{\vec{\ell}} e^{i\vec{\ell}\cdot \vec{x}} M_{L}^{\mu\nu}
  \int_{\vec{k}} \frac{f(\vec{k}) (k-\ell)_{L}}{|\vec{k}|^{n}}
\label{}\end{equation}
This integral is dimensionless and vanishes in dimensional regularization. This demonstrates that the source term only contributes to the 1PM order. 

In the case of the Kerr metric, the metric fluctuation always takes the form given in \eqref{conditionforField}. Moreover, as shown above, the Kerr source fully satisfies the locality condition. One may also explicitly verify that its Fourier transform contains no propagator as we have derived in \eqref{00sourceMomentum} and \eqref{0isourceMomentum}. Thus it is sufficient to solve \eqref{graviton_current_recursion} for deriving the Kerr metric.


\section{Series Expansion of the Kerr Metric in the Harmonic Gauge}\label{Appendix:C}
The explicit form of the metric perturbations $\mathfrak{h}^{\mu\nu}$ of the Kerr metric in the harmonic coordinates \eqref{KerrHarmonicCoordinate} in terms of the Landau-Lifshitz variable $\mathfrak{g}^{\mu\nu} = \eta^{\mu\nu} - \mathfrak{h}^{\mu\nu}$ around a flat background is as follows:
\begin{equation}
\begin{aligned}
  \mathfrak{h}^{00}
  &=
  \frac{G M R^2 \left(a^2 (3 G M+4 R)+(G M+R) \left(G^2 M^2+3 G M R+4 R^2\right)\right)}{\left(a^2 z^2+R^4\right) \left(a^2-G^2 M^2+R^2\right)}\,,
  \\
  \mathfrak{h}^{01}
  &=
  -\frac{2 a G M R^2 y (G M+R)}{\left(a^2 z^2+R^4\right) \left(a^2-G^2 M^2+R^2\right)}\,,
  \\
  \mathfrak{h}^{02}
  &=
  \frac{2 a G M R^2 x (G M+R)}{\left(a^2 z^2+R^4\right) \left(a^2-G^2 M^2+R^2\right)}\,,
  \\
  \mathfrak{h}^{03}
  &= 0\,,
  \\
  \mathfrak{h}^{11}
  &=
  \frac{G^2 M^2 R^2 (a y+R x)^2}{\left(a^2+R^2\right)^2 \left(a^2 z^2+R^4\right)}\,,
  \qquad
  \mathfrak{h}^{12}
  =
  \frac{G^2 M^2 R^2 (a y+R x) (R y-a x)}{\left(a^2+R^2\right)^2 \left(a^2 z^2+R^4\right)}\,,
  \\
  \mathfrak{h}^{13}
  &=
  \frac{G^2 M^2 R z (a y+R x)}{\left(a^2+R^2\right) \left(a^2 z^2+R^4\right)}\,,
  \qquad\ \,
  \mathfrak{h}^{22}
  =
  \frac{G^2 M^2 R^2 (a x-R y)^2}{\left(a^2+R^2\right)^2 \left(a^2 z^2+R^4\right)}\,,
  \\
  \mathfrak{h}^{23}
  &=
  \frac{G^2 M^2 R z (R y-a x)}{\left(a^2+R^2\right) \left(a^2 z^2+R^4\right)}\,,
  \qquad\ \,
  \mathfrak{h}^{33}
  =
  \frac{G^2 M^2 z^2}{a^2 z^2+R^4}\,.
\end{aligned}\label{}
\end{equation}

We further expand the above $\mathfrak{h}^{\mu\nu}$ with respect to the small $G$ and $a$ in the Cartesian coordinates. In this case, $R$ satisfies the defining equation, $\frac{x^{2}+y^{2}}{r^{2}+a^{2}}+\frac{z^{2}}{r^{2}}=1$, and it is represented by 
\begin{equation}
  R = \sqrt{\frac{r^{2}-a^{2}+\sqrt{(r^{2}-a^{2})^{2}+4a^{2}z^{2}}}{2}}
\label{}\end{equation}
where $r^{2} = x^{2}+y^{2}+z^{2}$.

The general form of the double expansion of each component can be written in terms of the Legendre polynomial $P_{n} (x)$, and Jacobi polynomials $P^{(\alpha,\beta)}_{n} (x)$,
\begin{equation}
\begin{aligned}
  \mathfrak{h}^{00}\big|_{1,2n}
  &=
  \frac{4(-1)^{n}}{r^{2n+1}} P_{2n}\Big(\frac{z}{r}\Big)\,,
  &\qquad
  \mathfrak{h}^{00}\big|_{1,2n+1} &= 0\,,
  \\
  \mathfrak{h}^{00}\big|_{2,2n}
  &= 
  \frac{G^{2} M^{2} a^{2 n}}{r^{2 n+2}}\left[3 P_{n}(u)+4 P_{n}^{(-1,1)}(u)\right] \,,
  &\qquad
  \mathfrak{h}^{00}\big|_{2,2n+1} &= 0\,,
  \\
  \mathfrak{h}^{00}\big|_{2m+1,2n}
  &=
  \frac{8(-1)^{n} \left(\frac{1}{2}\right)_{n}}{\left(m+\frac{1}{2}\right)_{n}} \frac{ M^{2 m+1}}{r^{2m+2 n+1}} P_{n}^{\left(-\frac{1}{2}, m\right)}\left(u\right)\,,
  &\qquad
  \mathfrak{h}^{00}\big|_{2m+1,2n+1}
  &= 0 \,,
  \\
  \mathfrak{h}^{00}\big|_{2m,2n}
  &= 
  \frac{4M^{2 m}}{r^{2 m+2 n}}\left[P_{n}^{(0, m-1)}\left(u\right)+P_{n}^{(-1, m)}\left(u\right)\right]\,,
  &\qquad
  \mathfrak{h}^{00}\big|_{2m,2n+1}
  &= 0\,,
\end{aligned}\label{Perturbation_h00}
\end{equation}
where $u=1-\frac{2z^2}{r^{2}}$, $m\geq1$ and $n \geq 0$ .
Next, we consider the off-diagonal block, $\mathfrak{h}^{0 i}$. 
\begin{equation}
\begin{aligned}
  \mathfrak{h}^{01}\big|_{2m,2n+1}
  &=
  - \frac{2M^{2m} y}{r^{2m+2 n+2}} P_{n}^{(0, m)}\left(u\right)\,,
  &\qquad
  \mathfrak{h}^{01}\big|_{2m,2n}
  &= 0\,,
  \\
  \mathfrak{h}^{01}\big|_{2m+1,2n+1}
  &=
  -\frac{2 M^{2 m+1} y}{r^{2 p+2 n+3}} P_{n}^{\left(-\frac{1}{2}, m+1\right)}\left(u\right)  \,,
  &\qquad
  \mathfrak{h}^{01}\big|_{2m+1,2n}
  &= 0\,,
  \\
  \mathfrak{h}^{02}\big|_{m,n}
  &=
  \mathfrak{h}^{01}\big|_{m,n}\Big|_{x \to -y\,,~y \to x} \,,
  &\qquad
  \mathfrak{h}^{03}\big|_{m,n} &= 0\,.
\end{aligned}\label{Perturbation_h0i}
\end{equation}

Finally, let us consider the spatial component block. First, the even sector is given by
\begin{equation}
\begin{aligned}
  \mathfrak{h}^{11}\big|_{2,2n}
  &=
  \frac{M^{2}}{r^{2 n+4}}\left[x^{2} P_{n}^{(-1,2)}(u)+y^{2} P_{n-1}^{(0,2)}(u)\right] \,,
  \\
  \mathfrak{h}^{22}\big|_{2,2n}
  &=
  \mathfrak{h}^{11}\big|_{2,2n} \Big|_{x\to y\,,~ y\to x} \,,
  \\
  \mathfrak{h}^{33}\big|_{2,2n}
  &=
  \mathfrak{h}^{11}\big|_{2,2n} \Big|_{x\to z\,,~ y\to 0} \,,
  \\
  \mathfrak{h}^{12}\big|_{2,2n}
  &=
  \frac{M^{2} x y}{r^{2 n+4}}\left[P_{n}^{(-1,2)}(u)-P_{n-1}^{(0,2)}(u)\right]\,,
  \\
  \mathfrak{h}^{13}\big|_{2,2n}
  &=
  \frac{M^{2} x z}{r^{2 n+4}} P_{n}^{(0,1)}(u)
  \\
  \mathfrak{h}^{23}\big|_{2,2n}
  &=
  \mathfrak{h}^{13}\big|_{2,2n} \Big|_{x\to y} \,,
\end{aligned}\label{}
\end{equation}
Next, the odd sector, identified as a pure gauge, is 
\begin{equation}
\begin{aligned}
  \mathfrak{h}^{11}\big|_{2,2n+1}
  &=
  \frac{2M^{2} x y}{r^{2 n+5}} P_{n}^{\left(-\frac{1}{2}, 2\right)}(u)\,,
  \\
  \mathfrak{h}^{22}\big|_{2,2n+1}
  &=
  - \mathfrak{h}^{11}\big|_{2,2n+1} \,,
  \\
  \mathfrak{h}^{33}\big|_{2,2n+1}
  &= 0 \,,
  \\
  \mathfrak{h}^{12}\big|_{2,2n+1}
  &=
  -\frac{M^{2}\left(x^{2}-y^{2}\right)}{r^{2 n+5}} P_{n}^{\left(-\frac{1}{2}, 2\right)}(u),\,,
  \\
  \mathfrak{h}^{13}\big|_{2,2n-1}
  &=
  \frac{M^{2} y z}{r^{2 n+5}} P_{n}^{\left(\frac{1}{2}, 1\right)}(u)\,,
  \\
  \mathfrak{h}^{23}\big|_{2,2n-1}
  &=
  - \mathfrak{h}^{13}\big|_{2,2n-1}\Big|_{y\to x}
\end{aligned}\label{}
\end{equation}
%


\section{Explicit Expressions for the Currents}

In this appendix, we show all the currents

\subsection{2PM}\label{Appendix:D.1}
We first present the general expression $\mathfrak{J}^{00}|^{2,n}_{\ell}$ and $\mathfrak{J}^{0i}|^{2,n}_{\ell}$:
\begin{equation}
\begin{aligned}
  \mathfrak{J}^{00}\big|^{2,2n}_{\ell}
  &=
  \frac{(-1)^{n}M^{2} \pi^{\frac{3}{2}-\epsilon} \Gamma\left(\frac{1}{2}-\epsilon-n\right)}{2^{3 n+2 \epsilon-1}(n!)^{2}|\boldsymbol{\ell}|^{1-2\epsilon}} 
  \\&\quad
  \times\sum_{k=0}^{n} (-1)^{k}\binom{n}{k}(6 k-7)(2 k-3)!!\,
  (\ell_{3})^{2n-2k} |\boldsymbol{\ell}|^{2k}\,
  \prod_{j=k}^{n-1}(2 \epsilon+2 j+1) 
  \\
  \mathfrak{J}^{0i}\big|^{2,2n+1}_{\ell}
  &=
  \frac{i(-1)^{n} M^{2} \pi^{\frac{3}{2}-\epsilon} \Gamma\left(\frac{1}{2}-\epsilon-n\right)}{2^{3 n+2 \epsilon-1} n!(n+1)!} \frac{\epsilon^{ij3}\ell_j}{|\boldsymbol\ell|^{1-2\epsilon}}
  \\&\quad\times
  \sum_{k=0}^{n}(-1)^{k}\binom{n}{k}(2 k-1)!!\,
  (\ell_{3})^{2n-2k} |\boldsymbol{\ell}|^{2k}\,
  \prod_{j=k}^{n-1}(2 \epsilon+2 j+1)
\end{aligned}\label{}
\end{equation}
The
\begin{equation}
\begin{aligned}
  \mathfrak{J}^{00}\big|^{2,2n}_{\ell}
  &=
  \frac{7 M^2\pi^{\frac{3}{2}-\epsilon}\,\Gamma (\frac{1}{2}-\epsilon)}{2^{2\epsilon+2n-1}(n!)^2}
  \frac{(\ell_3)^{2n}}{|\boldsymbol\ell|^{1-2\epsilon}}\
  {}_3F_2\!\left[\begin{matrix}-n,\,-\frac12,\,-\frac16\\-\frac76,\,\frac12+\epsilon\end{matrix};\frac{|\boldsymbol\ell|^{2}}{\ell_3^2}\right]\,,
  \\
  \mathfrak{J}^{0i}\big|^{2,2n+1}_{\ell}
  &=
  \frac{iM^2\pi^{\frac{3}{2}-\epsilon}\, \Gamma (\frac{1}{2}-\epsilon)}{2^{2\epsilon+2n-1}n!(n+1)!}
  \frac{\epsilon^{ij3}\ell_j(\ell_3)^{2n}}{|\boldsymbol\ell|^{1-2\epsilon}}\,
  {}_2F_1\!\left[ \begin{matrix} -n,\frac12\\ \frac12+\epsilon \end{matrix}; \frac{|\boldsymbol\ell|^2}{\ell_3^2} \right]\,,
\end{aligned}\label{}
\end{equation}
where $_3F_2$ and $_2F_1$ are hypergeometric functions. The first few terms are given by
\begin{equation}
\begin{aligned}
\mathfrak{J}^{00}\big|^{2}_{\ell}
  &=
  \frac{7M^2\pi^{\frac{3}{2}-\epsilon}\,\Gamma\!\left(\tfrac{1}{2}-\epsilon\right)}{2^{2\epsilon-1}\, |\boldsymbol{\ell}|^{1-2\epsilon}}
  -a^2 \frac{M^2\, \pi^{\frac{3}{2}-\epsilon}\,\Gamma\!\left(-\epsilon-\tfrac{1}{2}\right)}{2^{2\epsilon+2}|\boldsymbol{\ell}|^{1-2\epsilon}} 
  \Bigl[\,|\boldsymbol{\ell}|^2 + 7\ell_3^2\,(2\epsilon+1)\Bigr]
  \\& \quad
  +a^4 \frac{M^2 \pi^{\frac{3}{2}-\epsilon}\,\Gamma\!\left(-\epsilon{-}\tfrac{3}{2}\right)}{2^{7+2\epsilon}|\boldsymbol{\ell}|^{1-2\epsilon}}
  \Bigl[\,7\ell_3^4(4\epsilon^{2}{+}8\epsilon{+}3)+2\ell_3^2(2\epsilon+3)\,|\boldsymbol{\ell}|^2
  + 5|\boldsymbol{\ell}|^4\Bigr]
  + \mathcal{O}(a^{6})\,,
  \\
  \mathfrak{J}^{0i}\big|^{2}_{\ell}
  &=
  i\frac{M^2\, \pi^{\frac{3}{2}-\epsilon}}{2^{2\epsilon-1}|\boldsymbol{\ell}|^{1-2\epsilon}} \epsilon^{ij3}\ell_j
  \bigg[a\, \Gamma\!\left(\tfrac{1}{2}{-}\epsilon\right)-\frac{a^{3}}{2^{4}}\Gamma\!\left(-\tfrac{1}{2}-\epsilon\right)\Bigl((1+2\epsilon)\,\ell_3^2 - |\boldsymbol{\ell}|^2\Bigr)\bigg]
  + \mathcal{O}(a^{5})\,.
\end{aligned}\label{}
\end{equation}

Next, we consider the diagonal spatial components, $\mathfrak{J}^{ii}|^{2,n}_{\ell}$. Solving the 2PM recursion yields contributions only from the even power in $a$
\begin{equation}
\begin{aligned}
  \mathfrak{J}^{11}\big|^{2,2n}_{\ell}
  &=
  \frac{(-1)^{n} M^{2} \pi^{\frac{3}{2}-\epsilon} \Gamma\left(\frac{1}{2}-\epsilon-n\right)}{2^{3 n+2\epsilon} n!(n+1)!\left|\boldsymbol{\ell}\right|^{3-2\epsilon}}
  \mathcal{P}_{n}\,,
  \\
  \mathfrak{J}^{22}\big|^{2,2n}_{\ell}
  &=
  \mathfrak{J}^{11}\big|^{2,2n}_{\ell} \Big|_{\ell_{1}\leftrightarrow \ell_{2}}\,,
  \\
  \mathfrak{J}^{33}\big|^{2,2n}_{\ell}
  &=
  \mathfrak{J}^{11}\big|^{2,2n}_{\ell} \Big|_{\ell_{1}\to \ell_{3}, \ell_{2}\to0}\,,
\end{aligned}\label{}
\end{equation}
where
\begin{equation}
\begin{aligned}
  \mathcal{P}_{n} 
  &= 
  \ell_{1}^{2}\prod_{j=0}^{n}(2 \epsilon+2 j-1)
  -\sum_{k=1}^{n+1}(-1)^{k}(2 k{-}3)!! \,
  \binom{n}{k{-}1} \ell_{3}^{2n-2k+2} |\boldsymbol{\ell}|^{2k}
  \prod_{j=k}^{n}(2 \epsilon+2 j-1)
  \\&\quad
  + \sum_{k=1}^{n}(-1)^{k-1}(2 k-3)!!\, (\ell_{1}^{2}-2 k \ell_{2}^{2}) \ell_{3}^{2n-2k} |\boldsymbol{\ell}|^{2k} \binom{n}{k}\prod_{j=k}^{n}(2 \epsilon+2 j-1)\,, 
  \\
\end{aligned}\label{}
\end{equation}
Using the hypergeometric function, $\mathfrak{J}^{11}|^{2,2n}_{\ell}$ reduces to
\begin{equation}
\begin{aligned}
  \mathfrak{J}^{11}\big|^{2,2n}_{\ell}
  &=
  \frac{M^2\pi^{\frac{3}{2}-\epsilon}\,\Gamma\!\left(\frac12-\epsilon\right)}{2^{2\epsilon+2n-1}n!(n+1)!} \frac{\ell_3^{2n}}{|\boldsymbol{\ell}|^{3-2\epsilon}}
  \Bigg[
  \big(\epsilon-\tfrac12\big) \ell_1^2\, {}_2F_1\!\left[\begin{matrix}-n,\,-\frac12\\\epsilon-\frac12 \end{matrix}; \frac{\left|\boldsymbol{\ell}\right|^2}{\ell_3^2} \right]
  \\&\qquad\qquad\qquad\qquad
  +\frac{\left|\boldsymbol{\ell}\right|^2}{2} {}_2F_1\!\left[\begin{matrix}-n,\,\frac12\\\epsilon+\frac12\end{matrix};\frac{\left|\boldsymbol{\ell}\right|^2}{\ell_3^2}\right]
  - \frac{n\ell_2^2\left|\boldsymbol{\ell}\right|^2}{\ell_3^2}
      {}_2F_1\!\left[\begin{matrix}1-n,\,\frac12\\\epsilon+\frac12\end{matrix};\frac{\left|\boldsymbol{\ell}\right|^2}{\ell_3^2}\right]
  \Bigg] \,.
\end{aligned}\label{}
\end{equation}
Their first few terms of $\mathfrak{J}^{11}|^{2,2n}_{\ell}$ are
\begin{equation}
\begin{aligned}
  \mathfrak{J}^{11}\big|^{2}_{\ell}
  &=
  \frac{M^2 \pi^{\frac{3}{2}-\epsilon} \Gamma\!\left(\tfrac{1}{2}-\epsilon\right)}{4^{\epsilon}|\boldsymbol{\ell}|^{3-2\epsilon}}
  \Bigl((2\epsilon-1)\,\ell_1^2 + |\boldsymbol{\ell}|^2\Bigr)
  \\&\quad+
  a^{2} \frac{M^2 \pi^{\frac{3}{2}-\epsilon} \Gamma\!\left(-\tfrac{1}{2}-\epsilon\right)}{4^{2+\epsilon}|\boldsymbol{\ell}|^{3-2\epsilon}}
  \Bigl[(1{-}4\epsilon^2)\,\ell_1^2\,\ell_3^2
  - (1{+}2\epsilon)(\ell_1^2 - 2\ell_2^2 + \ell_3^2)\,|\boldsymbol{\ell}|^2
  + |\boldsymbol{\ell}|^4\Bigr]
  \\&\quad
  + a^{4} \frac{M^2\, \pi^{\frac{3}{2}-\epsilon}\,\Gamma\!\left(-\tfrac{3}{2}-\epsilon\right)}{3\times 4^{4+\epsilon}|\boldsymbol{\ell}|^{3-2\epsilon}} 
  \Bigl[
  3\,|\boldsymbol{\ell}|^6 + \bigl(3+4\epsilon(2+\epsilon)\bigr)\,\ell_3^2\,(2\ell_1^2-4\ell_2^2+\ell_3^2)\,|\boldsymbol{\ell}|^2
  \\&\qquad\qquad
  + (2\epsilon{-}1)(2\epsilon{+}1)(2\epsilon{+}3)\,\ell_1^2\,\ell_3^4
  - (3{+}2\epsilon)(\ell_1^2{-}4\ell_2^2{+}2\ell_3^2)\,|\boldsymbol{\ell}|^4
  \Bigr]
  + \mathcal{O}(a^{6})\,.
\end{aligned}\label{}
\end{equation}

Finally, we focus on the spatial off-diagonal components, $\mathfrak{J}^{12}|^{2,2n}_{\ell}$, $\mathfrak{J}^{13}|^{2,2n}_{\ell}$ and $\mathfrak{J}^{23}|^{2,2n}_{\ell}$ 
\begin{equation}
\begin{aligned}
  \mathfrak{J}^{12}\big|^{2,2n}_{\ell}
  &=
  \frac{(-1)^{n+1}}{n!(n+1)!}\frac{M^{2} \pi^{\frac{3}{2} -\epsilon} \Gamma\left(\frac{3}{2}-\epsilon-n\right)}{2^{3 n+2 \epsilon-1}} \frac{\ell_{1} \ell_{2}}{|\boldsymbol{\ell}|^{3-2\epsilon}}
  \\&\quad\times
  \sum_{k=0}^{n}\binom{n}{k} (-1)^{k-1}(2 k+1)(2 k-3)!! \,
  (\ell_{3})^{2n-2k} |\boldsymbol{\ell}|^{2k} \prod_{j=k}^{n-1}(2 \epsilon+2 j-1) \,,
  \\
  \mathfrak{J}^{13}\big|^{2,2n}_{\ell}
  &=
  \frac{(-1)^{n+1}}{n!(n+1)!}\frac{M^{2} \pi^{\frac{3}{2} -\epsilon} \Gamma\left(\frac{3}{2}-\epsilon-n\right)}{2^{3 n+2 \epsilon-1}} \frac{\ell_{1} \ell_{3}}{|\boldsymbol{\ell}|^{3-2\epsilon}}
  \\&\quad\times
  \sum_{k=0}^{n}\binom{n}{k} (-1)^{k-1}(2 k-3)!!\, 
  (\ell_{3})^{2n-2k} |\boldsymbol{\ell}|^{2k}
  \prod_{j=k}^{n-1}\big(2 \epsilon+2 j-1\big)\,.
\end{aligned}\label{}
\end{equation}
We may represent the series expansion in terms of the following compact form:
\begin{equation}
\begin{aligned}
  \mathfrak{J}^{12}\big|^{2,2n}_{\ell}
  &=
  - \frac{M^2\pi^{3/2-\epsilon}
  \Gamma\!\left(\frac32-\epsilon\right)}{2^{2\epsilon+2n-1} n!(n+1)!} 
  \frac{\ell_1\ell_2\ell_3^{2n}}{\left|\boldsymbol{\ell}\right|^{3-2\epsilon}}\,
  {}_3F_2\!\left[ \begin{matrix} -n,\,-\frac12,\,\frac32\\\frac12,\,\epsilon-\frac12 \end{matrix} ;\frac{\left|\boldsymbol{\ell}\right|^2}{\ell_3^2}\right]\,,
  \\
  \mathfrak{J}^{13}\big|^{2,2n}_{\ell} 
  &=
  - \frac{M^2\pi^{3/2-\epsilon} \Gamma\!\left(\frac32-\epsilon\right)}{2^{2\epsilon+2n-1} n!(n+1)!}
  \frac{\ell_1\ell_3^{2n+1}}{\left|\boldsymbol{\ell}\right|^{3-2\epsilon}}\,
  {}_2F_1\!\left[\begin{matrix}-n,\,-\frac12\\\epsilon-\frac12\end{matrix};\frac{\left|\boldsymbol{\ell}\right|^2}{\ell_3^2}\right]\,,
  \\
  \mathfrak{J}^{23}\big|^{2,2n}_{\ell} 
  &=
  \mathfrak{J}^{13}_{(2,2n)|\ell} \Big|_{\ell_{1}\to \ell_{2}}\,.
\end{aligned}\label{}
\end{equation}
Their first few terms are
\begin{equation}
\begin{aligned}
  \mathfrak{J}^{12}\big|^{2}_{\ell} 
  &=
  - \frac{M^2 \pi^{\frac{3}{2}-\epsilon}\, \Gamma\!\left(\tfrac{3}{2}-\epsilon\right)\ell_1 \ell_2}{2^{-1+2\epsilon}|\boldsymbol{\ell}|^{3-2\epsilon}} 
  + a^2\frac{M^2\, \pi^{\frac{3}{2}-\epsilon} \Gamma\!\left(\tfrac{1}{2}-\epsilon\right)\ell_1 \ell_2}{2^{3+2\epsilon}|\boldsymbol{\ell}|^{3-2\epsilon}} 
  \Big[(2\epsilon-1)\,\ell_3^2 + 3\,|\boldsymbol{\ell}|^2\Big]
  \\&\quad
  +a^4\frac{M^2\, \pi^{\frac{3}{2}-\epsilon}\, \Gamma\!\left({-}\tfrac{1}{2}{-}\epsilon\right)\ell_1 \ell_2}{3\times 2^{7+2\epsilon}|\boldsymbol{\ell}|^{3-2\epsilon}}
  \Bigl[(1{-}4\epsilon^2)\ell_3^4
  - 6(1+2\epsilon)\ell_3^2|\boldsymbol{\ell}|^2
  + 5|\boldsymbol{\ell}|^4\Bigr]
  + \mathcal{O}(a^{6})\,,
  \\
  \mathfrak{J}^{13}\big|^{2}_{\ell} 
  &=
  - \frac{M^2 \pi^{\frac{3}{2}-\epsilon}\,
  \Gamma\!\left(\tfrac{3}{2}-\epsilon\right)\ell_1\ell_3}{2^{-1+2\epsilon}|\boldsymbol{\ell}|^{3-2\epsilon}}
  +
  a^{2}\frac{M^2\, \pi^{\frac{3}{2}-\epsilon}\, \Gamma\!\left(\tfrac{1}{2}-\epsilon\right)\ell_1\ell_3}{2^{-3-2\epsilon}|\boldsymbol{\ell}|^{3-2\epsilon}}
  \Big[(2\epsilon-1)\,\ell_3^2 + |\boldsymbol{\ell}|^2\Big]
  \\&\quad
  + a^{4} \frac{M^2\, \pi^{\frac{3}{2}-\epsilon}\, \Gamma\!\left({-}\tfrac{1}{2}{-}\epsilon\right) \ell_1\ell_3}{3\times 2^{7+2\epsilon}|\boldsymbol{\ell}|^{3-2\epsilon}}
  \Bigl[(1-4\epsilon^2)\,\ell_3^4
  - 2(1+2\epsilon)\ell_3^2|\boldsymbol{\ell}|^2
  + |\boldsymbol{\ell}|^4\Bigr]
  + \mathcal{O}(a^{6})\,.
\end{aligned}\label{}
\end{equation}
%

\subsection{4PM order}\label{Appendix:D.2}

\begin{equation}
\begin{aligned}
  \mathfrak{J}^{00} \big|^{4,2n}_{\ell}
  &=
  \frac{M^{4} \pi^{\frac{3}{2}-\epsilon} \Gamma\!\left(-\frac{1}{2}-\epsilon-n\right) }{2^{3 n+2 \epsilon-2} n!(n+1)!} |\boldsymbol{\ell}|^{1+2\epsilon} 
  \\&\quad\times
  \sum_{k=0}^{n}\binom{n}{k} (-1)^{n+k} (k-1)(2 k-3)!! (\ell_{3})^{2n-2k} |\boldsymbol{\ell}|^{2k} \prod_{j=k+1}^{n}(2 \epsilon+2 j+1)\,,
  \\
  \mathfrak{J}^{0i} \big|^{4,2n+1}_{\ell}
  &=
  i \frac{M^{4} \pi^{\frac{3}{2}-\epsilon} \Gamma\left(-\frac{1}{2}-\epsilon-n\right)}{2^{1+3 n+2 \epsilon}n!(n+2)!}  \epsilon^{ij3}\ell_{j} |\boldsymbol{\ell}|^{1+2\epsilon}
  \\&\quad
  \times
  \sum_{k=0}^{n}(-1)^{n-k}\binom{n}{k}(2 k-1)!!\,
  q^{n-k} s^{k}\,
  \prod_{j=k+1}^{n}(2 \epsilon+2 j+1)\,.
\end{aligned}\label{}
\end{equation}
These can be written in terms of the hypergeometric functions
\begin{equation}
\begin{aligned}
  \mathfrak{J}^{00} \big|^{4,2n}_{\ell}
  &=
  \frac{M^{4} \pi^{\frac{3}{2}-\epsilon} \Gamma\!\left(-\frac{1}{2}-\epsilon\right)}{2^{2n+2 \epsilon-2}n!(n+1)!} |\boldsymbol{\ell}|^{1+2\epsilon} (\ell_{3})^{2 n} 
  \\& 
  \times\left[{ }_{2} F_{1}\bigg[\begin{array}{c}-n,-\frac{1}{2} \\ \epsilon+\frac{3}{2}\end{array} ; \tfrac{|\boldsymbol{\ell}|^{2}}{\ell_{3}^{2}}\bigg]
  -\frac{n |\boldsymbol{\ell}|^{2}}{(2 \epsilon+3) \ell_{3}^{2}}\,{}_{2} F_{1}\bigg[\begin{array}{c}1-n, \frac{1}{2} \\ \epsilon+\frac{5}{2}\end{array} ; \tfrac{|\boldsymbol{\ell}|^{2}}{\ell_{3}^{2}}\bigg]\right]\,,
  \\
  \mathfrak{J}^{0i} \big|^{4,2n+1}_{\ell}
  &=
  i \frac{M^{4} \pi^{\frac{3}{2}-\epsilon} \Gamma\left(-\frac{1}{2}-\epsilon\right)}{2^{1+2n+2 \epsilon}n!(n+2)!} 
  \epsilon^{ij3}\ell_{j} |\boldsymbol{\ell}|^{1+2\epsilon} (\ell_{3})^{2 n}\,
  {}_{2} F_{1}\bigg[\begin{array}{l}-n, \frac{1}{2} \\ \epsilon+\frac{3}{2}\end{array} ; \tfrac{|\boldsymbol{\ell}|^{2}}{\ell_{3}^{2}}\bigg]\,,
\end{aligned}\label{}
\end{equation}
This result is completely consistent with the all-order metric in \eqref{Perturbation_h00} and \eqref{Perturbation_h0i}.

Next let us consider the case in which the 2PM gauge terms are absent. We denote the currents as $\mathfrak{J}'^{\mu\nu}|_{4,n}$. Again, we read off the following general results:
\begin{equation}
\begin{aligned}
  \mathfrak{J}'^{00} \big|^{4,2n}_{\ell}
  &=
  \frac{M^{4} \pi^{\frac{3}{2}-\epsilon} \Gamma\left(-\frac{3}{2}-\epsilon\right)}{2^{2n+2 \epsilon} n!(n+1)!} |\boldsymbol{\ell}|^{1+2\epsilon}
  \\&\quad
  \times \sum_{k=0}^{n} \frac{(-n)_{k}}{\left(\epsilon+\frac{5}{2}\right)_{k} k!}\left[\frac{8 n+17}{2(n+2)}\left(-\tfrac{1}{2}\right)_{k}+\frac{3}{4}\left(\tfrac{1}{2}\right)_{k}+\left(\tfrac{3}{2}\right)_{k}\right] (\ell_{3})^{2n-2k} |\boldsymbol{\ell}|^{2k}\,,
  \\
  \mathfrak{J}'^{0i} \big|^{4,2n}_{\ell}
  &=
  -i \frac{3M^{4} \pi^{\frac{3}{2} -\epsilon} \Gamma\!\left(-\frac{3}{2}-\epsilon\right) }{2^{2 n+2+2 \epsilon}n!(n+2)!} \epsilon^{ij3}\ell_{j}|\boldsymbol{\ell}|^{1+2\epsilon}
  \sum_{k=0}^{n} \frac{(-n)_{k}\left(\frac{1}{2}\right)_{k}\left(\frac{5}{2}\right)_{k}}{\left(\frac{3}{2}\right)_{k}\left(\epsilon+\frac{5}{2}\right)_{k} k!} \ell_{3}^{2n-2k} |\boldsymbol{\ell}|^{2k}\,.
\end{aligned}\label{}
\end{equation}
We may reduce the expansion in terms of the hypergeometric functions
\begin{equation}
\begin{aligned}
  \mathfrak{J}'^{00} \big|^{4,2n}_{\ell}
  &=
  \frac{M^{4} \pi^{\frac{3}{2}-\epsilon} \Gamma\!\left(-\frac{3}{2}-\epsilon\right)}{2^{2n+2 \epsilon}n!(n+1)!} |\boldsymbol{\ell}|^{1+2\epsilon} (\ell_{3})^{2n} 
  \Bigg[~
    \frac{8 n{+}17}{2n{+}4}\, {}_{2} F_{1}\bigg[\begin{array}{c}-n,-\frac{1}{2} \\ \epsilon+\frac{5}{2}\end{array} ; \tfrac{|\boldsymbol{\ell}|^{2}}{\ell_{3}^{2}}\bigg]
  \\&\qquad\qquad\qquad\qquad\qquad\qquad
  + \frac{3}{4}{ }_{2} F_{1}\bigg[\begin{array}{c}-n, \frac{1}{2} \\ \epsilon+\frac{5}{2}\end{array} ; \tfrac{|\boldsymbol{\ell}|^{2}}{\ell_{3}^{2}}\bigg]
  + {}_{2} F_{1}\bigg[\begin{array}{l}-n, \frac{3}{2} \\ \epsilon+\frac{5}{2}\end{array} ; \tfrac{|\boldsymbol{\ell}|^{2}}{\ell_{3}^{2}}\bigg]~\Bigg]\,,
  \\
  \mathfrak{J}'^{0i} \big|^{4,2n+1}_{\ell}
  &=
  -i \frac{3M^4 \pi^{\frac{3}{2}-\epsilon} \Gamma\!\left(-\frac32-\epsilon\right)}{2^{2n+2+2\epsilon}n!(n+2)!} \epsilon^{ij3}\ell_j |\boldsymbol{\ell}|^{1+2\epsilon} \ell_3^{2n} \,
  { }_{3} F_{2}\bigg[\begin{array}{l}-n, \frac{1}{2}, \frac{5}{2} \\ \frac{3}{2}, \epsilon+\frac{5}{2}\end{array} ; \frac{|\boldsymbol{\ell}|^{2}}{\ell_{3}^{2}}\bigg]\,,
\end{aligned}\label{}
\end{equation}

Next, we consider the spatial diagonal components:
\begin{equation}
\begin{aligned}
  \mathfrak{J}'^{11} \big|^{4,2n}_{\ell}
  &=
  -\frac{M^{4} \pi^{\frac{3}{2}-\epsilon} \Gamma\left(-\frac{3}{2}-\epsilon\right)}{2^{2n+2\epsilon+2}(n-1)!(n+2)!} 
  \left|\boldsymbol{\ell}\right|^{1+2\epsilon}
  \\&\quad\times
  \Bigg[ \frac{2\epsilon+3}{3} \sum_{k=0}^{n-1} \frac{(1-n)_k\left(\frac12\right)_k\left(\frac32\right)_k}{\left(\frac52\right)_k
  \left(\epsilon+\frac32\right)_k k!} \left(\ell_1^2-2(k+2)\ell_2^2\right) \ell_3^{2(n-1-k)} \left|\boldsymbol{\ell}\right|^{2k}
  \\&\hspace{1.5cm}
  - \sum_{k=0}^{n-1} \frac{(1-n)_k\left(\frac12\right)_k}{\left(\epsilon+\frac52\right)_k k!} \ell_3^{2(n-1-k)} \left|\boldsymbol{\ell}\right|^{2k+2}
  \Bigg],
  \qquad n\geq 1 .
  \\
  \mathfrak{J}'^{22} \big|^{4,2n}_{\ell}
  &=
  \mathfrak{J}'^{11} \big|^{4,2n}_{\ell}\Big|_{\ell_{1}\leftrightarrow \ell_{2}}\,,
  \\
  \mathfrak{J}'^{33} \big|^{4,2n}_{\ell}
  &=
  - \frac{M^4\pi^{\frac{3}{2}-\epsilon} \Gamma\!\left(-\frac32-\epsilon\right)}{2^{2n+1+2\epsilon}n!(n+2)!}\, \left|\boldsymbol{\ell}\right|^{1+2\epsilon}\ell_3^{2n}
  \\&\quad
  \times
  \Bigg[\frac{n+2}{2}\,{}_2F_1\!\left[\begin{matrix}-n,\frac12\\\epsilon+\frac52\end{matrix};\frac{\left|\boldsymbol{\ell}\right|^2}{\ell_3^2}\right]
  - {}_2F_1\!\left[\begin{matrix} -n,-\frac12\\\epsilon+\frac52\end{matrix};\frac{\left|\boldsymbol{\ell}\right|^2}{\ell_3^2}\right]
  \Bigg]\,, \qquad n\ge1
\end{aligned}\label{}
\end{equation}
These are reduced to the closed formula
\begin{equation}
\begin{aligned}
  \mathfrak{J}'^{11} \big|^{4,2n}_{\ell}
  &=
  \frac{M^{4} \pi^{\frac{3}{2}-\epsilon} \Gamma\!\left(-\frac{3}{2}-\epsilon\right)}{2^{2n+2\epsilon+2}(n-1)!(n+2)!} \left|\boldsymbol{\ell}\right|^{1+2\epsilon}\ell_3^{2n-2}
  \\&\quad\times
  \Bigg[
  \frac{2(1-n)}{5}\,\frac{\ell_2^2\left|\boldsymbol{\ell}\right|^2}{\ell_3^2}{}_3F_2\!\left[\begin{matrix}2-n,\frac32,\frac52\\\frac72,\epsilon+\frac52\end{matrix};\frac{\left|\boldsymbol{\ell}\right|^2}{\ell_3^2}\right]
  + \left|\boldsymbol{\ell}\right|^2{}_2F_1\!\left[\begin{matrix}1-n,\frac12\\\epsilon+\frac52\end{matrix};\frac{\left|\boldsymbol{\ell}\right|^2}{\ell_3^2}\right]
  \\&\hspace{1.cm}
  -\frac{2\epsilon+3}{3}\left(\ell_1^2-4\ell_2^2\right){}_3F_2\!\left[\begin{matrix}1-n,\frac12,\frac32\\\frac52,\epsilon+\frac32\end{matrix};\frac{\left|\boldsymbol{\ell}\right|^2}{\ell_3^2}\right]
\Bigg],
  \\
  \mathfrak{J}'^{33} \big|^{4,2n}_{\ell}
  &=
  - \frac{M^4\pi^{\frac{3}{2}-\epsilon} \Gamma\!\left(-\frac32-\epsilon\right)}{2^{2n+1+2\epsilon}n!(n+2)!}\, \left|\boldsymbol{\ell}\right|^{1+2\epsilon}\ell_3^{2n}
  \\&\quad
  \times
  \Bigg[\frac{n+2}{2}\,{}_2F_1\!\left[\begin{matrix}-n,\frac12\\\epsilon+\frac52\end{matrix};\frac{\left|\boldsymbol{\ell}\right|^2}{\ell_3^2}\right]
  - {}_2F_1\!\left[\begin{matrix} -n,-\frac12\\\epsilon+\frac52\end{matrix};\frac{\left|\boldsymbol{\ell}\right|^2}{\ell_3^2}\right]
  \Bigg]\,.
\end{aligned}\label{}
\end{equation}

Finally, the general expression of the currents of the off-diagonal spatial parts is
\begin{equation}
\begin{aligned}
  \mathfrak{J}'^{12} \big|^{4,2n}_{\ell}
  &=
  -\frac{M^{4} \pi^{\frac{3}{2}-\epsilon} \Gamma\left(-\frac{3}{2}-\epsilon\right)\ell_{1} \ell_{2}}{2^{2n+2\epsilon+2}(n-1)!(n+2)!} 
  \sum_{k=0}^{n-1} \frac{(1{-}n)_{k}\left[4\left(\frac{1}{2}\right)_{k}{+}\left(\frac{3}{2}\right)_{k}\right]}{\left(\epsilon+\frac{5}{2}\right)_{k} k!} \frac{\left|\boldsymbol{\ell}\right|^{2k+1+2\epsilon}}{\ell_{3}^{2k+2-2n}}
  \,,
  \\
  \mathfrak{J}'^{13} \big|^{4,2n}_{\ell}
  &=
  -\frac{M^{4} \pi^{\frac{2}{}-\epsilon} \Gamma\left(-\frac{3}{2}-\epsilon\right)\ell_{1} \ell_{3}}{2^{2n+2\epsilon+2}(n-1)!(n+2)!} 
  \sum_{k=0}^{n-1} \frac{(1-n)_{k}\left(\frac{1}{2}\right)_{k}}{\left(\epsilon+\frac{5}{2}\right)_{k} k!}  \frac{\left|\boldsymbol{\ell}\right|^{2k+1+2\epsilon}}{\ell_{3}^{2k+2-2n}}\,,
  \\
  \mathfrak{J}'^{23} \big|^{4,2n}_{\ell}
  &=
  \mathfrak{J}'^{13} \big|^{4,2n}_{\ell}\Big|_{\ell_{1}\to \ell_{2}}
\end{aligned}\label{}
\end{equation}
where $n\geq 1$. The currents vanish for $n=0$ case, $\mathfrak{J}'^{12}|^{4,0}_{\ell} = \mathfrak{J}'^{13}|^{4,0}_{\ell} = \mathfrak{J}'^{23}|^{4,0}_{\ell} = 0$. 
Their closed form is
\begin{equation}
\begin{aligned}
  \mathfrak{J}'^{12} \big|^{4,2n}_{\ell}
  &=
  - \frac{5\,M^4 \pi^{\frac{3}{2}-\epsilon} \Gamma\!\left(-\frac32-\epsilon\right)}{2^{2n+2\epsilon+2}(n-1)!(n+2)!}\, \ell_1\ell_2(\ell_3)^{2n-2}\, \left|\boldsymbol{\ell}\right|^{1+2\epsilon}
  {}_3F_2\!\left[\begin{matrix}1-n,\frac12,\frac72\\\frac52,\epsilon+\frac52\end{matrix}; \frac{\left|\boldsymbol{\ell}\right|^2}{\ell_3^2} \right]\,,
  \\
  \mathfrak{J}'^{13} \big|^{4,2n}_{\ell}
  &=
  - \frac{M^4\pi^{\frac{3}{2}-\epsilon}\Gamma\!\left(-\frac32-\epsilon\right)}{2^{2n+2\epsilon+2} (n-1)!(n+2)!}\,
  \ell_1\ell_3^{2n-1}\left|\boldsymbol{\ell}\right|^{1+2\epsilon}\,
  {}_2F_1\!\left[\begin{matrix}1-n,\frac12\\\epsilon+\frac52\end{matrix} ;
\frac{\left|\boldsymbol{\ell}\right|^2}{\ell_3^2} \right]\,,
\end{aligned}\label{}
\end{equation}

Taking inverse Fourier transformation, we obtain the general $a^{n}$-th order terms for the 4PM metric perturbations, denoted as $\mathfrak{h}'^{\mu\nu}|_{4,n}$. First, $\mathfrak{h}'^{00}|_{4,n}$ and $\mathfrak{h}'^{0i}|_{4,n}$ are
\begin{equation}
\begin{aligned}
  \mathfrak{h}'^{00}\big|_{4,n}
  &=
  \frac{M^{4}}{r^{2n+4}}\bigg[
  \frac{23 n+48}{3(n+2)} P_{n}^{(0,1)}(u)-\frac{59n+126}{15(n+2)} P_{n-1}^{(0,2)}(u)
  \\&\qquad\qquad\quad
  +\frac{2 n+3}{n+2} \sum_{j=2}^{n} \frac{2^{j-2} j!}{(2 j+3)!!} P_{n-j}^{(0, j+1)}(u)
  \bigg]\,,
  \\
  \mathfrak{h}'^{0i}\big|_{4,n}
  &=
  -2 \frac{a^{2 n+1} G^{4} M^{4} \epsilon^{ij3}x_{j}}{r^{2 n+6}} P_{n}^{(0,2)}(u)\,,
\end{aligned}\label{}
\end{equation}
where $u=1- 2z^2/r^{2}$ and $P^{(\alpha,\beta)}_{n} (u)$ is the Jacobi polynomial. 

Next, we consider the spatial components
\begin{equation}
\begin{aligned}
  \mathfrak{h}'^{ij}|_{4,0} 
  &= 0\,,
  \\
  \mathfrak{h}'^{11}\big|_{4,2n}
  &=
  -\frac{ M^{4}}{r^{2n +8}}
  \left[
      x^{2} \bigg(\frac{1}{3} P_{m}^{(0,3)}(u) + F_{m}\bigg) 
    - y^{2} \bigg(\frac{4}{3} P_{m}^{(0,3)}(u) - F_{m}\bigg)
  \right]\,,
  \\
  \mathfrak{h}'^{22}\big|_{4,2n}
  &=
  \mathfrak{h}'^{11}\big|_{4,2n}\Big|_{x\leftrightarrow y} \,,
  \\
  \mathfrak{h}'^{33}\big|_{4,2n}
  &=
  \frac{M^4}{r^{2n+4}} \Bigg[ \frac{nP_{n}^{(0,1)}(u)}{3n+6}
  + \frac{(6-n)P_{n-1}^{(0,2)}(u)}{15\left(n+2\right)} 
  - \frac{2n+3}{n+2} \sum_{j=2}^{n} \frac{2^{j-2}j!P_{n-j}^{(0,j+1)}(u)}{(2j+3)!!}  \Bigg]\,,
  \\
  \mathfrak{h}'^{12}\big|_{4,2n}
  &=
  \frac{M^{4} x y}{r^{n+6}}\left[
    -\frac{5}{3} P_{\frac{n}{2}-1}^{(0,3)}(u)
    +\sum_{j=2}^{n / 2} \frac{2^{j-1}(j-1)!}{(2 j+1)!!} P_{\frac{n}{2}-j}^{(0, j+2)}( u)\right]\,,
  \\
  \mathfrak{h}'^{13}\big|_{4,2n}
  &=
  -\frac{M^{4} x z}{r^{2n+6}} \sum_{j=1}^{n} \frac{2^{j-1} j!}{(2 j+1)!!} P_{n-j}^{(0, j+2)}(u)\,,
  \\
  \mathfrak{h}'^{23}\big|_{4,2n}
  &=
  \mathfrak{h}'^{13}\big|_{4,2n}\Big|_{x\to y}\,,
\end{aligned}\label{}
\end{equation}
where $n\geq 1$ and $F_{m} =  \sum_{j=1}^{m} \frac{2^{j-1} j!}{(2 j+3)!!} P_{m-j}^{(0, j+3)}(u)$.


\end{document}